\documentclass[a4paper,aps,prl,reprint,showpacs,floatfix]{revtex4-2}
\usepackage{amsmath, amssymb}
\usepackage{bm}
\usepackage{multirow}
\usepackage{braket}
\usepackage{graphicx}
\usepackage{booktabs}
\usepackage[unicode=true,colorlinks=true,linkcolor=blue,citecolor=blue,urlcolor=blue]{hyperref}
\usepackage{xspace}
\usepackage{newtxtext}
\usepackage{mathrsfs}
\usepackage[english]{babel}
\usepackage[utf8]{inputenc}

\newcommand{\rr}[1]{#1} 

\newcommand{\tc}{$T_{{\rm c}}$\xspace}
\newcommand{\cpb}{CsPbBr$_{3}$\xspace}

\begin{document}
\title{First-Principles Phonon Quasiparticle Theory Applied to a Strongly Anharmonic Halide Perovskite}
\author{Terumasa Tadano}
\email{TADANO.Terumasa@nims.go.jp}
\affiliation{Research Center for Magnetic and Spintronic Materials, National Institute for Materials Science, Tsukuba 305-0047, Japan}

\author{Wissam A. Saidi}
\affiliation{Department of Mechanical Engineering \& Materials Science (MEMS), University of Pittsburgh, USA}

\date{\today}
\begin{abstract}
Understanding and predicting lattice dynamics in strongly anharmonic crystals is one of the long-standing challenges in condensed matter physics.
Here we propose a first-principles method that gives accurate quasiparticle (QP) peaks of the phonon spectrum with strong anharmonic broadening. On top of the conventional first-order self-consistent phonon (SC1) dynamical matrix, the proposed method incorporates frequency renormalization effects by the bubble self-energy within the QP approximation. We apply the developed methodology to the strongly anharmonic $\alpha$-CsPbBr$_3$ that displays phonon instability within the harmonic approximation in the whole Brillouin zone. While the SC1 theory significantly underestimates the cubic-to-tetragonal phase transition temperature (\tc) by more than 50\%, we show that our approach yields \tc = 404--423~K, in excellent agreement with the experimental value of 403~K. We also demonstrate that an accurate determination of QP peaks is paramount for quantitative prediction and elucidation of \rr{phonon linewidth}.
\end{abstract}
\maketitle

Lattice vibrations in functional materials often exhibit strong anharmonicity; i.e., thermal or quantum fluctuation of atoms is so large that lattice dynamics cannot be predicted accurately by the quasiharmonic phonon theory. Notable examples of such materials include perovskites~\cite{Kozina_NatPhys2019, Knoop_PRM2020, Haque_AdvSci2020}, thermoelectric materials~\cite{Delaire_NatureMaterials2011,Li_NaturePhysics2015,Suekuni_AdvMater2018}, and superconducting hydrides~\cite{Errea_PRL2015,Errea_Nature2020}. In particular, halide perovskites have been attracting growing interest due to their unique physical properties, including high photovoltaic performance~\cite{Huang_NatReviewMater2017} along with relevant electron-phonon coupled physics, and ultralow thermal conductivity~\cite{Haque_AdvSci2020,Xie_JACS2020}. However, an in-depth theoretical understanding and quantitative predictions of the lattice dynamics and phonon-related properties in these materials are still challenging due to the lack of first-principles computational approaches that can describe with high fidelity the intricate complexities associated with anharmonic behavior. 
In principle, lattice anharmonicity can be fully captured using \textit{ab initio} molecular dynamics (aiMD) based on density functional theory (DFT).
However, this approach is of limited use because it invariably requires the use of large supercells to capture phonon-phonon interactions involving nonzero-wavevector phonons and a long simulation time to extract well-converged values of the band- and momentum-resolved phonon frequencies and linewidths. Thus, such simulations can quickly develop into a computational bottleneck. 

\rr{To mitigate these challenges, several quasiparticle (QP)-like approaches have been proposed in the last decade~\cite{Hellman_PRB2011,Errea_PRB2014,Tadano_PRB2015,Roekeghem_PRB2016,Ravichandran_PRB2018}. Although QP approximation cannot describe satellite peaks, i.e., incoherent parts of spectra, it has several advantages. Namely, it simplifies the evaluation of physical quantities that can be directly compared with experiments, such as group velocity and heat capacity. Besides, the QP treatment gives an \textit{effective} one-body Hamiltonian of interacting phonons that is necessary as input for calculations of electron-phonon and phonon-phonon couplings in functional materials.}
The first-order self-consistent phonon (SC1) theory is one of the most successful methods, which determines the renormalized phonon frequencies by the variational principle applied to the first-order cumulant expansion of the Helmholtz free energy~\cite{Klein_JLTP1972, Tadano_JPSJ2018}. Since the SC1 theory can, to a large extent, remedy the negative frequency problems of the harmonic approximation, it has been actively employed in first-principles calculations of phonon-related physics of anharmonic materials, including thermal transport~\cite{Tadano_PRB2015,Kang_NanoLett2019,Xia_PRL2020,Xia_PRX2020,Kawano_JPCC2021}, phonon-limited mobility~\cite{Zhao_PRB2020}, bandgap renormalization~\cite{Patrick_PRB2015,Wu_JPCL2020}, thermal expansion~\cite{Oba_PRM2019,Kwon_PRB2020}, and conventional superconductivity~\cite{Errea_PRB2014,Errea_PRL2015,Sano_PRB2016}.

While these studies clearly demonstrate the advantage of the SC1 theory over the quasiharmonic theory and purely perturbative approaches, the quantitative accuracy of SC1 is still inadequate for strongly anharmonic materials. More specifically,  SC1 theory tends to overpredict phonon frequencies at finite temperatures because it neglects the frequency shift associated with the bubble self-energy. Indeed, as we will show below, such a shift is substantial in the strongly anharmonic \cpb and have significant effects on the theoretical phase transition temperature and lattice thermal conductivity (LTC).

In this Letter, we propose a first-principles phonon calculation method that gives accurate QP peaks of phonon spectrum broadened by phonon-phonon interactions. The developed method\rr{, which we formulate using the modern language employed in the $GW$ approximation in electronic structure theory~\cite{martin2016interacting},} incorporates the frequency shift by the bubble self-energy within the QP approximation and thereby solves the overestimation problem inherent to the SC1 theory. We apply the developed method to cubic \cpb ($\alpha$ phase), which displays strong lattice anharmonicity accompanying the cubic-to-tetragonal phase transition at $T_{\mathrm{c}}=403$ K~\cite{Hirotsu_JPSJ1974}. Although the SC1 theory underpredicts \tc by more than 50\%, the QP theory gives \tc values of 404--423 K, which successfully reproduce the experimental value. We also show that the LTC of $\alpha$-\cpb calculated based on the QP theory combined with a beyond-Boltzmann treatment~\cite{Simoncelli_NatPhys2019} is ultralow ($<$ 0.5 W/mK at 500 K) and shows weak temperature dependence, whereas the LTC based on SC1 shows a clear trend of overestimation, thus highlighting the importance of an accurate determination of QP peaks for quantitative prediction of \rr{phonon linewidth and LTC}.

Reliable modeling of lattice dynamics requires an accurate treatment of lattice anharmonicity, which is manifested as an interaction between non-interacting (harmonic) phonons. This problem can be formulated by the Dyson equation as 
\begin{equation}
    \{G_{q}(\omega)\}^{-1} = \{G_{q}^{0}(\omega)\}^{-1} - \Sigma_{q}[G](\omega),
    \label{eq:Dyson}
\end{equation}
where  $G_{\bm{q}jj'}^{0}(\omega)$ is the non-interacting phonon propagator and $\Sigma_{q}[G](\omega)$ is the anharmonic self-energy. For the self-energy, the most important terms associated with the third- and fourth-order anharmonicity are usually considered as $\Sigma_q[G](\omega) = \Sigma_q^{\mathrm{T}}[G, \Phi_3]+\Sigma_q^{\mathrm{L}}[G, \Phi_4]+\Sigma_q^{\mathrm{B}}[G,\Phi_3](\omega)$. Here, ``T'', ``L'', and ``B'' stand for the tadpole, loop, and bubble diagrams, respectively. Their dependence on anharmonic force constants $(\Phi_{3},\Phi_{4})$ is indicated explicitly. Once the above Dyson equation is solved for $G(\omega)$, the information of lattice dynamics can be obtained from the spectral function $A_q(\omega)=|\mathrm{Im}G_{q}(\omega)|/\pi$. However, achieving a fully self-consistent solution to Eq.~(\ref{eq:Dyson}) is challenging because both sides of the equation depend on $G$ and $\omega$.

The SC1 theory greatly simplifies Eq.~(\ref{eq:Dyson}) as
\begin{equation}
    \{G_{q}^{\mathrm{S}}(\omega)\}^{-1} = \{G_{q}^{0}(\omega)\}^{-1} - 
    \Sigma_q^{\mathrm{T}}[G^{\mathrm{S}}, \Phi_3]-\Sigma_q^{\mathrm{L}}[G^{\mathrm{S}}, \Phi_4], \label{eq:SC1}
\end{equation}
where the $\omega$-dependent bubble self-energy is dropped. 
While the SC1 theory is powerful in its versatility and reasonable accuracy,
the frequency shift associated with the neglected bubble diagram is not small~\cite{Tadano_JPSJ2018,Errea_Nature2020} and particularly large in $\alpha$-\cpb, as we will show below. Hence, $\Sigma_q^{\mathrm{B}}[G,\Phi_3](\omega)$ should be included. Given that the SC1 propagator $G^{\mathrm{S}}_q(\omega)$ is reasonably close to the fully-dressed propagator $G_q(\omega)$, we may simplify Eq.~(\ref{eq:Dyson}) as
\begin{equation}
    \{G_{q}(\omega)\}^{-1} \approx \{G^{\mathrm{S}}_q(\omega)\}^{-1} -\Sigma_q^{\mathrm{B}}[G^{\mathrm{S}},\Phi_3](\omega),
    \label{eq:bubble_pert}
\end{equation}
where the self-consistency for $G$ is lifted. 
This is similar to the $G_0W_0$ approximation, where the Kohn--Sham wavefunction is used for the non-interacting part, and the correlation is treated in ``one-shot'' with $\Sigma=iG_{0}W_{0}$. So far, Eq.~(\ref{eq:bubble_pert}) has been employed to calculate the phonon spectral function of anharmonic solids~\cite{Tadano_JPSJ2018,Aseginolaza_PRL2019}.
However, instead of calculating the $\omega$-dependent propagator, we aim to develop an \textit{effective} one-body Hamiltonian that well represents the QP peaks given by Eq.~(\ref{eq:bubble_pert}). To this end, we propose the following self-consistent equation~\cite{Note1}:
\begin{equation}
    \Omega_{\bm{q}\nu}^2 = (\omega_{\bm{q}\nu}^{\mathrm{S}})^{2} - 2\omega_{\bm{q}\nu}^{\mathrm{S}}\mathrm{Re}\Sigma_{\bm{q}\nu}^{\mathrm{B}}[G^{\mathrm{S}},\Phi_3](\omega=\Omega_{\bm{q}\nu}). \label{eq:QP}
\end{equation}
Here, $\omega_{\bm{q}\nu}^{\mathrm{S}}$ is the SC1 frequency, and the bubble self-energy is evaluated at the QP frequency $\Omega_{\bm{q}\nu}$. The above nonlinear equation, which again resembles the QP approximation used in the $GW$ calculations~\cite{martin2016interacting}, needs to be solved self-consistently with respect to $\Omega_{\bm{q}\nu}$. To simplify this procedure, it is tempting to linearize  $\Sigma_{\bm{q}\nu}^{\mathrm{B}}[G^{\mathrm{S}},\Phi_3](\Omega_{\bm{q}\nu})$ around $\Omega_{\bm{q}\nu}=\omega_{\bm{q}\nu}^{\mathrm{S}}$, which yields $\Omega_{\bm{q}\nu}^2 = (\omega_{\bm{q}\nu}^{\mathrm{S}})^{2} - 2Z_{\bm{q}\nu}\omega_{\bm{q}\nu}^{\mathrm{S}}\mathrm{Re}\Sigma_{\bm{q}\nu}^{\mathrm{B}}[G^{\mathrm{S}},\Phi_3](\omega_{\bm{q}\nu}^{\mathrm{S}})$ with $Z_{\bm{q}\nu}=[1+(\partial \mathrm{Re}\Sigma_{\bm{q}\nu}^{\mathrm{B}}/\partial \omega)|_{\omega=\omega_{\bm{q}\nu}^{\mathrm{S}}}]^{-1}$ being the renormalization factor. However, we found that this linearization yields a non-smooth temperature dependence of $\Omega_{\bm{q}\nu}$ due to the complex $\omega$ dependence of $\mathrm{Re}\Sigma_{\bm{q}\nu}^{\mathrm{B}}(\omega)$. Hence, we do not employ the linearization in this study. Instead of the nonlinear QP equation defined by Eq.~(\ref{eq:QP}), which we call QP-NL, several different QP treatments are possible. The simplest approximation is the static approximation ($\omega=0$), which incorporates the first-order correction term that appears in the Hessian of the SC1 free energy~\cite{Bianco_PRB2017}. The second option is to set $\omega=\omega_{\bm{q}\nu}^{\mathrm{S}}$. For the clarity of the following discussion, we denote these QP methods as QP[0] and QP[S], respectively. QP[S] is equivalent to setting $Z_{\bm{q}\nu}=1$ in the above linearized equation. 
We note that beyond this ``one-shot'' treatment of the bubble self-energy would be possible with an approximation akin to that used in the QP self-consistent $GW$ method~\cite{Schilfgaarde_PRL2006}, which is left for a future study.

\begin{figure}
    \centering
    \includegraphics[width=0.48\textwidth]{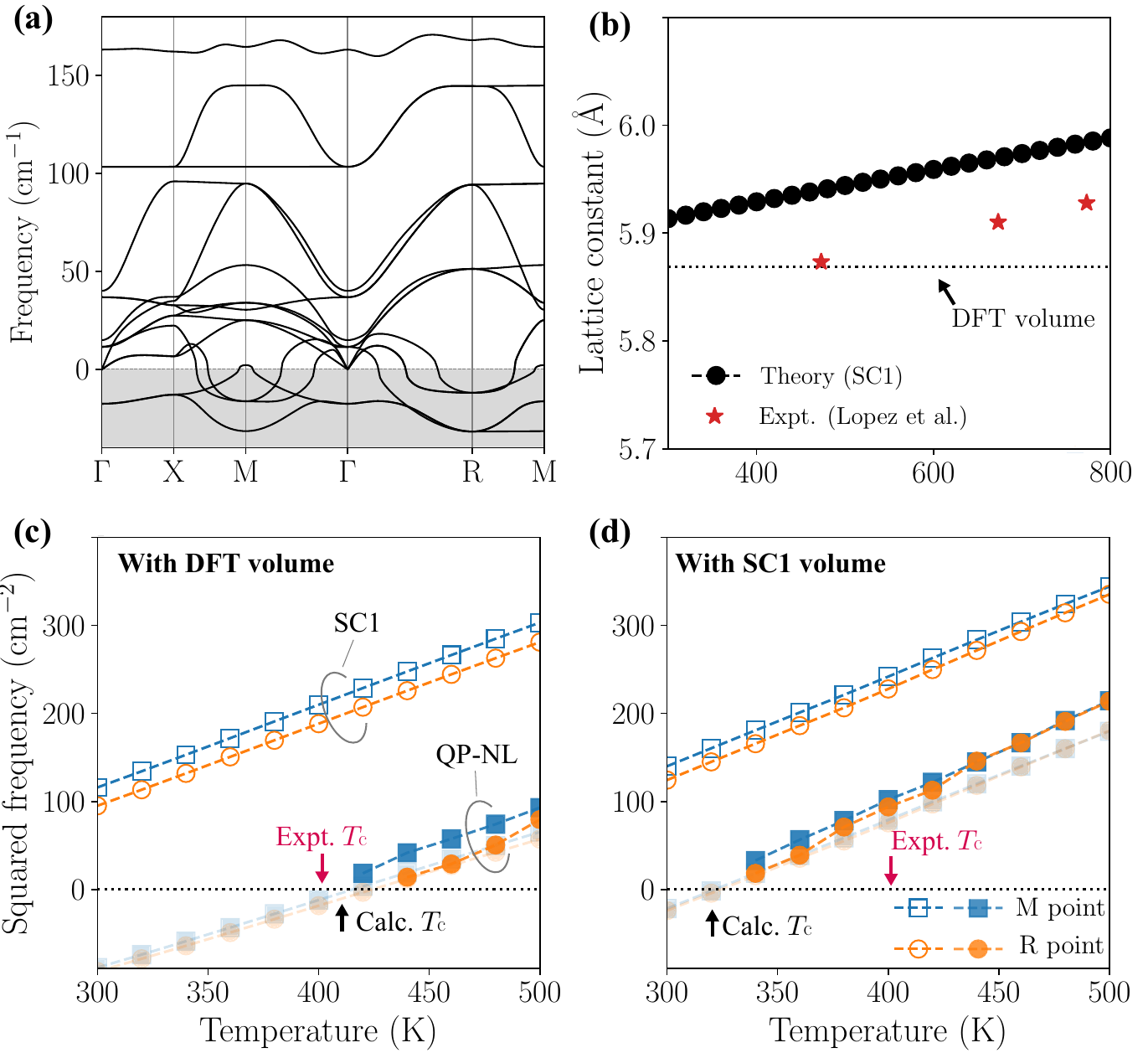}
    \caption{Calculated phonon frequency and lattice constant of $\alpha$-\cpb. (a) Harmonic phonon dispersion calculated with $V=V_{0}$. (b) Temperature-dependence of $V^{(S)}(T)$ compared with the experimental data of López \textit{et al.}~\cite{Lopez_Omega2020} (c,d) Temperature-dependence of the squared frequency of soft modes at M and R points with $V=V_{0}$ and $V=V^{(\mathrm{S})}(T)$. \rr{The QP[0] results are shown by translucent symbols.}}
    \label{fig:lattice_param_and_softmode}
\end{figure}

We now apply the developed QP-NL method to strongly anharmonic $\alpha$-\cpb. The DFT calculations were conducted using \textsc{quantum ESPRESSO}~\cite{QE2017}, with the GGA-PBEsol functional~\cite{PBEsol}. The lattice dynamics calculations were performed using  \textsc{alamode}~\cite{Tadano2014}. To include thermal expansion effects, we evaluated $F_{\mathrm{vib}}^{\mathrm{(S)}}(V,T)$ at various volumes and temperatures. The $T$-dependent volume was then obtained by minimizing the Helmholtz free energy as $V^{\mathrm{(S)}}(T)=\arg\min_{V} \{E_{0}(V)+F_{\mathrm{vib}}^{\mathrm{(S)}}(V,T)\}$. This approach was shown to work even for the cases where the quasiharmonic theory breaks down due to the presence of unstable modes~\cite{Oba_PRM2019}.
Harmonic and anharmonic \rr{interatomic force constants} necessary for the present lattice dynamics calculations were estimated by using a compressed sensing approach~\cite{Zhou_PRL2014}, for which we employed the adaptive LASSO~\cite{adalasso}.
More details are provided in the Supplementary Material (SM)~\cite{supplement}.

We first discuss the temperature dependence of the lattice constant shown in  Fig.~\ref{fig:lattice_param_and_softmode}(b). The optimized value obtained by DFT is 5.868~\AA{} that agrees exceptionally well with the experimental value of 5.873~\AA{} at 473 K~\cite{Lopez_Omega2020}. However, this almost perfect agreement is accidental as inferred after accounting for phonon excitations. Namely, at the SC1 level, we obtained 5.941~\AA{} at 480 K that overestimates 
the experimental value by $\sim$ 1\%. The calculated linear thermal expansion coefficient of $\alpha\simeq 25\times 10^{-6}$ K$^{-1}$ is within the range of experimental results 28--33$\times 10^{-6}$ K$^{-1}$~\cite{Rodova_JTAC2003}.

In $\alpha$-\cpb, phonon softening occurs in the whole Brillouin zone, as can be inferred already at the harmonic level (Fig.~\ref{fig:lattice_param_and_softmode}(a)). 
After accounting for anharmonic effects, these soft modes became dynamically stable in the high-temperature region and their frequencies decrease gradually with cooling following the Curie--Weiss law, as we elaborate below. We observed that the lowest frequency occurs at R$(\frac{1}{2}, \frac{1}{2}, \frac{1}{2})$ point and the second-lowest at M$(\frac{1}{2}, \frac{1}{2}, 0)$ point; the calculated temperature dependence of these soft modes are shown in Figs.~\ref{fig:lattice_param_and_softmode}(c) and (d). The SC1 theory always yields stable phonons when the self-consistent equation [Eq.~(\ref{eq:SC1})] converges. Notwithstanding, we can estimate \tc by fitting the linear part of $(\omega_{\bm{q}\nu}^{\mathrm{S}})^{2}$ with $A(T-T_{\mathrm{c}})$, 
see Table~\ref{tab:critical_temperature}. 
Given that the cubic-to-tetragonal phase transition is first order~\cite{Hirotsu_JPSJ1974} with a small temperature hysteresis \rr{of $\sim$7 K~\cite{Svirskas_JMCA}}, the prediction based on the Curie--Weiss law should be interpreted as a lower bound of \tc. As seen from the table, SC1 significantly overestimates the soft modes frequencies and thereby underestimates \tc, which is less than 50\% of the experimental \tc of 403 K. 

\begin{table}[!bth]
    \centering
    \caption{Critical temperature (K) of the cubic-to-tetragonal phase transition calculated at different levels of the QP theory. Two values in each cell show the \tc values estimated from the soft mode frequency at M and R points, respectively. The experimental \tc is 403 K~\cite{Hirotsu_JPSJ1974}.}
    \label{tab:critical_temperature}
    \begin{ruledtabular}
    \begin{tabular}{lcc}
        Method &  DFT volume & \rr{SC1} volume \\ \hline
         SC1 (Eq.~(\ref{eq:SC1})) & 177, 198 & 164, 183 \\
         QP[0] & 415, 424 & 322, 324\\
         QP[S] & 369, 382 & 303, 307\\
         QP-NL (Eq.~(\ref{eq:QP})) & 404, 423 & 319, 324 \\
    \end{tabular}
    \end{ruledtabular}
\end{table}

Including the bubble diagram by solving Eq.~(\ref{eq:QP}), we can see from  Fig.~\ref{fig:lattice_param_and_softmode}(c) and Table \ref{tab:critical_temperature} that the \tc value is in better agreement with experiment.  The QP-NL method with $V=V_{0}$ gives \tc value of 404--423 K, which agrees reasonably well with the experimental value. Also, the static QP[0] theory led to \tc values that are similar to those of the QP-NL method. This is reasonable because the QP energy approaches zero ($\Omega_{\bm{q}\nu}\rightarrow 0$) in the limit of $T\rightarrow T_{\mathrm{c}}$; hence, $\Sigma_{\bm{q}\nu}^{\mathrm{B}}(\Omega_{\bm{q}\nu})$ approaches $\Sigma_{\bm{q}\nu}^{\mathrm{B}}(0)$. By contrast, the QP[S] frequencies were generally larger than that of the static approximation and \tc value became lower by $\sim$40 K. All of these results clearly highlight the significant effect of the bubble diagram. Moreover, we observed that the \tc value is quite sensitive to the lattice constant. When we used the SC1 volume, the frequencies of the soft modes at M and R points became larger than those obtained with the DFT volume, which can be attributed to their negative Gr\"{u}neisen parameters~\cite{supplement}. Consequently, the estimated \tc value decreases by $\sim$20\% even though the difference in the lattice constant is only $\sim$1\%\rr{ (see Sec.~S3 of the SM~\cite{supplement}). Since a DFT lattice constant depends on the choice of the exchange-correlation functional and pseudopotential, the present result indicates the importance of carefully choosing them in quantitative predictions of \tc for \cpb.  A similar sensitivity has also been reported for BaTiO$_{3}$~\cite{Ehsan_PRB2021}.}

\begin{figure}
    \centering
    \includegraphics[width=0.48\textwidth]{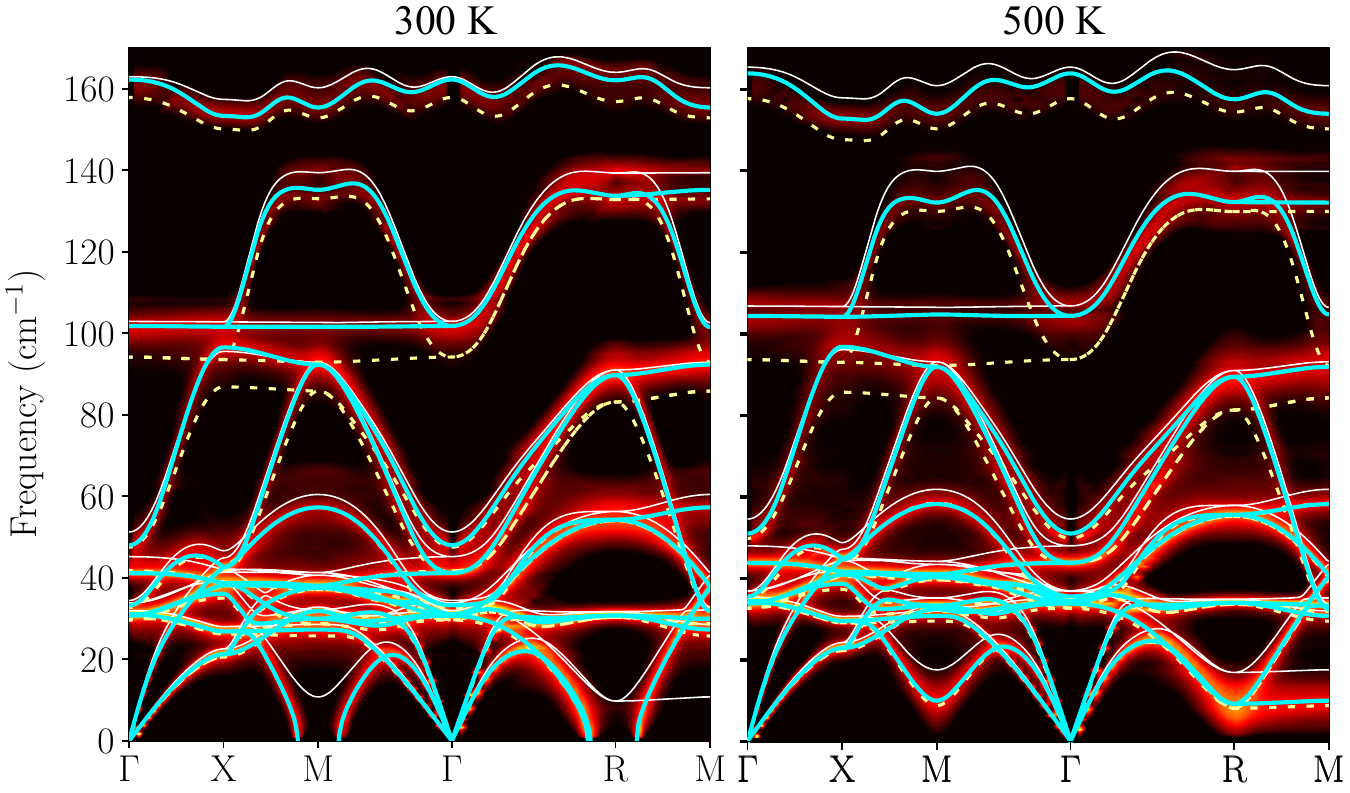}
    \caption{Anharmonic phonon dispersion curves and spectral function of cubic CsPbBr$_{3}$ calculated (a) below and (b) above $T_{\mathrm{c}}=403$ K. The white \rr{thin} lines, yellow \rr{dashed} lines, and cyan solid \rr{thick} lines represent the dispersion curves obtained within the SC1 theory, QP[0] theory, and the QP-NL theory, respectively. The colormap shows the spectral function $A_q(\omega)$. The volume is set to $V=V_0$.}
    \label{fig:phonon_dispersion}
\end{figure}


To obtain insight into the accuracy of the QP theory over a wider frequency range, the calculated anharmonic phonon dispersion curves are compared with the spectral function in Fig.~\ref{fig:phonon_dispersion}. Here, the spectral function $A_q(\omega)$ is obtained from $G_{q}(\omega)$ of Eq.~(\ref{eq:bubble_pert}) with full frequency-dependence of $\Sigma_{\bm{q}\nu}^{\mathrm{B}}(\omega)$ and is used as a reference to assess the accuracy of the QP theory. 
The SC1 theory tends to overestimate the phonon frequency as compared to the peak frequency of $A_q(\omega)$. The overestimation is particularly notable in the low-frequency soft modes that are still stable below \tc (left panel). The overestimation by the SC1 theory is mostly rectified by the QP theory, irrespective of the adopted value for $\omega$. However, Fig.~\ref{fig:phonon_dispersion} clearly shows that QP[0] underestimates the phonon frequencies above $\sim$70 cm$^{-1}$ as compared to $A_{q}(\omega)$. The underestimation of the optical \rr{modes} around 100 cm$^{-1}$ is as large as 10 cm$^{-1}$. 
We found that the QP-NL theory is free from such an under- or overestimation artifact and thereby best represents the peak frequency of $A_{q}(\omega)$, as shown by the solid lines in Fig.~\ref{fig:phonon_dispersion}. On the basis of these results, we posit that the QP-NL theory gives the most reliable \textit{effective} one-particle picture among the investigated approximations. We also found that the QP[S] gives similar results with the QP-NL although slightly overestimates the frequencies in the low-frequency region at low temperatures (see Fig.~S3 of the SM~\cite{supplement}). \rr{Although more comprehensive study is demanded, we expect the bubble frequency correction affects the soft-mode frequencies and associated physical properties, such as \tc and dielectric permittivity, in a broad range of materials that exhibit structural phase transition.}

\begin{figure}
    \centering
    \includegraphics[width=0.48\textwidth]{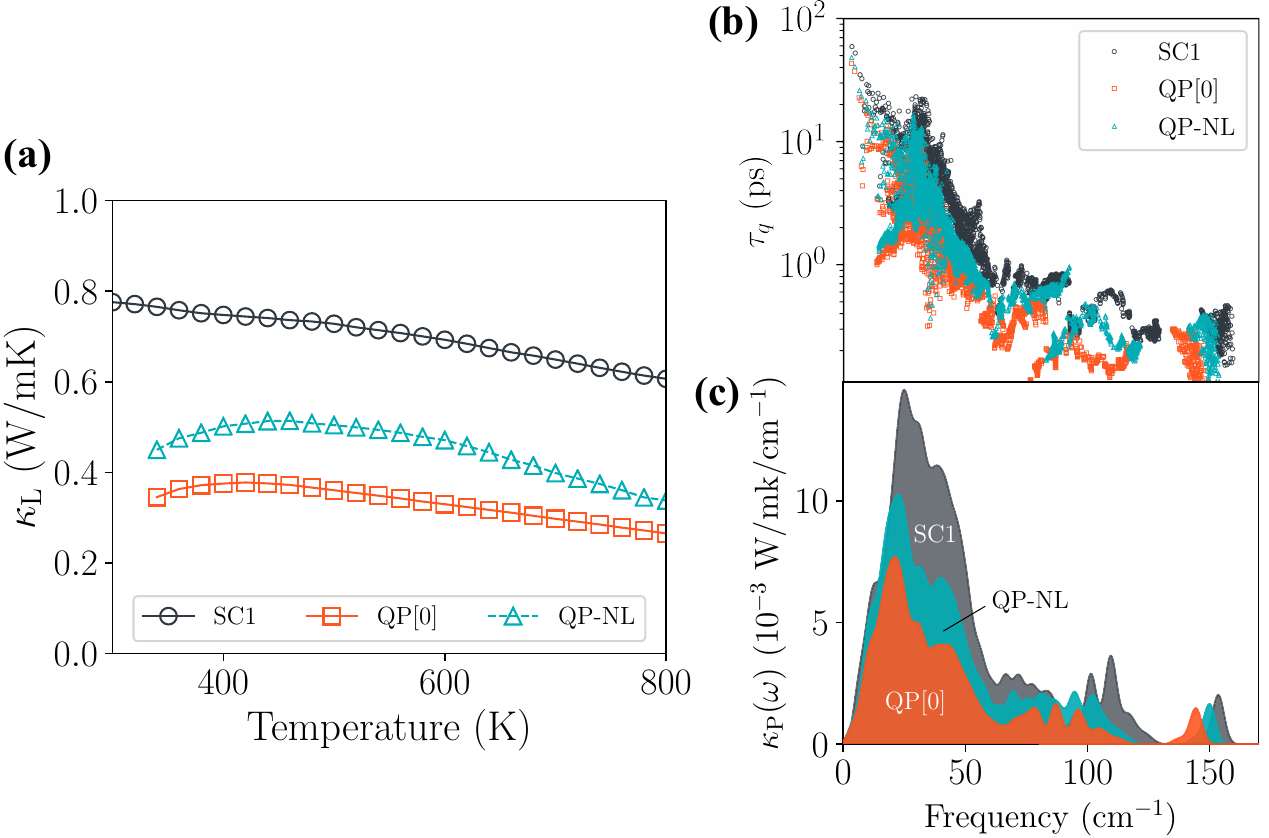}
    \caption{Lattice thermal conductivity and phonon lifetimes in $\alpha$-\cpb calculated using different dynamical matrices as inputs. (a) Lattice thermal conductivity $\kappa_{\mathrm{L}}$ [Eq.~(\ref{eq:kl})] above the theoretical \tc values. (b) Phonon lifetimes $\tau_{\bm{q}\nu}=\hbar/2\Gamma_{\bm{q}\nu}^{\mathrm{3ph}}$ at 500 K. (c) Spectral decomposition of the Peierls term calculated at 500 K. All calculations are done with $V=V_{\mathrm{S}}(T)$.}
    \label{fig:thermal_conductivity}
\end{figure}

The strong modifications of the phonon band structures due to anharmonic effects uncovered by the QP theories beyond SC1, particularly for the optical modes, is expected to have a strong influence on the \rr{phonon linewidth} of $\alpha$-\cpb. 
\rr{To see its influence quantitatively, we calculated the phonon linewidths as $\Gamma_{\bm{q}\nu}^{\mathrm{3ph}}=\mathrm{Im}\Sigma_{\bm{q}\nu}^{\mathrm{B}}[G,\Phi_3](\Omega_q)$. As shown in Figs.~\ref{fig:thermal_conductivity}(b), the difference in the input dynamical matrix results in notable change in the phonon lifetime (inverse linewidth); the lifetime becomes the longest (shortest) with the SC1 (QP[0]) frequency. When the phonon frequency is overestimated, the scattering phase space will be underestimated because of a smaller occupation number $n_{\bm{q}\nu}=[\exp{(\beta\hbar\omega_{\bm{q}\nu})}-1]^{-1}$. Besides, the strength of the three-phonon interaction will be smaller due to the weaker hybridization~\cite{Tadano_PRL2018}. These combined effects explain the factor two difference in $\tau_{\bm{q}\nu}$; with the SC1 frequency, the average phonon lifetime below 50 cm$^{-1}$ is $\sim$6.2 ps at 500 K, whereas it becomes $\sim$2.7 ps with the QP[0] frequency. 
}

\rr{
We compared the calculated phonon frequency and linewidth with the experimental values~\cite{Songvilay_PRM2019} for the transverse acoustic modes along the G-X and G-M lines. As shown in Fig.~S5 of the SM~\cite{supplement}, the SC1 overestimates the TA phonon frequencies, while QP[0] and QP-NL agree better with the experimental data. As for the linewidth, the calculated $\Gamma_{\bm{q}\nu}^{\mathrm{3ph}}$ was smaller than the experimental values even when the QP-NL dynamical matrix was used, which indicates the potential role of higher-order phonon scattering processes. To examine this, we also computed the four-phonon scattering rate $\Gamma_{\bm{q}\nu}^{\mathrm{4ph}}$ following Refs.~\cite{Tripathi_Cimento1974,Feng_PRB2017}. As shown in Fig.~S5, the total linewidth $\Gamma_{\bm{q}\nu}=\Gamma_{\bm{q}\nu}^{\mathrm{3ph}}+\Gamma_{\bm{q}\nu}^{\mathrm{4ph}}$ agrees reasonably well with the experimental values, but the agreement is observed only when the QP-NL dynamical matrix is used. 
}

\rr{Next, we investigate the influence of the input dynamical matrix on LTC of $\alpha$-\cpb. To this end, we evaluated the LTC using a two-channel model as~\cite{Simoncelli_NatPhys2019}}
\begin{multline}
 \kappa_{\mathrm{L}} = \frac{1}{N_qV}\sum_{\bm{q}\nu\nu'}\frac{c_{\bm{q}\nu}\omega_{\bm{q}\nu'} + c_{\bm{q}\nu'}\omega_{\bm{q}\nu} }{\omega_{\bm{q}\nu}+\omega_{\bm{q}\nu'}} \bm{v}_{\bm{q}\nu\nu'}\otimes \bm{v}_{\bm{q}\nu'\nu} \\
 \times \frac{\Gamma_{\bm{q}\nu}+\Gamma_{\bm{q}\nu'}}{(\omega_{\bm{q}\nu}-\omega_{\bm{q}\nu'})^2+(\Gamma_{\bm{q}\nu}+\Gamma_{\bm{q}\nu'})^2}, \label{eq:kl}
\end{multline}
where $c_{\bm{q}\nu}$ is the mode heat capacity, $V$ is the unit-cell volume, and $\bm{v}_{\bm{q}\nu\nu'}=\frac{1}{2}(\omega_{\bm{q}\nu}\omega_{\bm{q}\nu'})^{-\frac{1}{2}}\braket{\eta_{\bm{q}\nu}|\partial_{\bm{q}}C(\bm{q})|\eta_{\bm{q}\nu'}}$ is the inter-band generalization of the group velocity~\cite{Allen_PRB1993} with $C(\bm{q})$ and $\ket{\eta_{\bm{q}\nu}}$ being the dynamical matrix and polarization vector, respectively. The band diagonal term ($\nu=\nu'$) corresponds to the Peierls contribution ($\kappa_{\mathrm{P}}$) within the relaxation-time approximation, whereas the off-diagonal term gives the coherent contribution ($\kappa_{\mathrm{C}}$); the total LTC is given as $\kappa_{\mathrm{L}}=\kappa_{\mathrm{P}}+\kappa_{\mathrm{C}}$. When calculating the phonon frequency and $\bm{v}_{\bm{q}\nu\nu'}$, we used the \textit{effective} second-order force constants obtained either from the SC1 or QP eigenfrequencies/eigenvectors. 
\rr{Simoncelli \textit{et al.} applied Eq.~(\ref{eq:kl}) to orthorhombic \cpb, combined with $\Gamma_{\bm{q}\nu} \approx \Gamma_{\bm{q}\nu}^{\mathrm{3ph}}$, and obtained excellent agreements with experimental LTC~\cite{Simoncelli_NatPhys2019}. Hence, we employ the same approximation for $\Gamma_{\bm{q}\nu}$ in this study. Note that the inclusion of the four-phonon scattering process in Eq.~(\ref{eq:kl}) is technically straightforward, but it can underestimate LTC of strongly anharmonic materials because of the missing contributions from anharmonic heat flux~\cite{Sun_PRB2010}. 
}

Figure \ref{fig:thermal_conductivity}(a) shows the temperature-dependent $\kappa_{\mathrm{L}}$ calculated with three different dynamical matrices: SC1, QP[0], and QP-NL. As seen in the figure, SC1 gives the largest $\kappa_{\mathrm{L}}$ while QP[0] gives the smallest values. The difference mostly originated from $\kappa_{\mathrm{P}}$, and $\kappa_{\mathrm{C}}$ was rather insensitive to the adopted dynamical matrix. 
\rr{The calculated phonon lifetimes and the spectra of the Peierls term $\kappa_{\mathrm{P}}(\omega)$ are shown in Figs.~\ref{fig:thermal_conductivity}(b) and (c), respectively. The $\kappa_{\mathrm{P}}(\omega)$ data clearly shows that $\kappa_{\mathrm{P}}$ is dominated by}
the low-frequency phonons below 50 cm$^{-1}$, and the difference among SC1, QP[0], and QP-NL is most notable in this frequency region. From Fig.~\ref{fig:phonon_dispersion} (right panel), the phonon group velocity 
$\bm{v}_{\bm{q}\nu\nu}$ does not change appreciably in the three QP theories at 500 K. Indeed, the difference in $\kappa_{\mathrm{P}}$ can be attributed to the phonon lifetime.
Consequently, a factor two difference was observed also in $\kappa_{\mathrm{P}}$: 0.68, 0.30, and 0.44 W/mK with the SC1, QP[0], and QP-NL frequencies, respectively. Interestingly, we found that the phonon lifetimes \rr{$\tau_{\bm{q}\nu}=\hbar/2\Gamma_{\bm{q}\nu}^{\mathrm{3ph}}$} of $\alpha$-\cpb are as large as those in $\alpha$-SrTiO$_{3}$~\cite{Tadano_JPSJ2018}, whose LTC at 500 K is larger than 7 W/mK. \rr{Hence, the small group velocity also contributes to realizing ultralow LTC of $\alpha$-\cpb, in accord with the previous interpretation~\cite{Elbaz_NanoLetters2017}}.

For the coherent term, $\kappa_{\mathrm{C}}$, we obtained 0.051, 0.065, and 0.060 W/mK at 500 K using SC1, QP[0], and QP-NL dynamical matrices, respectively; these values were nearly temperature independent above \tc. In comparison to the $\kappa_{\mathrm{C}}$ value of 0.3 W/mK (300 K) reported for the orthorhombic \cpb~\cite{Simoncelli_NatPhys2019}, the coherent term for the cubic phase was smaller by a factor of $\sim$5--6. This is reasonable considering that the number of phonon branches is 15 in the cubic phase, while it is 60 in the orthorhombic phase. Even for the cubic phase, the coherent term accounts for more than 13\% of the total LTC when the QP dynamical matrix is used and therefore should not be neglected. The total LTC predicted by Eq.~(\ref{eq:kl}) shows a temperature dependence that is much weaker than $\kappa_{\mathrm{L}} \propto T^{-1}$. A similar weak $T$-dependence has been observed in previous experimental and computational studies of halide perovskites~\cite{Wang_AdvFuncMater2016,Xie_JACS2020} and other anharmonic solids~\cite{Tadano_JPSJ2018,Suekuni_AdvMater2018}, which can be mainly attributed to the temperature-induced hardening of optical modes.

The $\kappa_{\mathrm{L}}$ values at 500 K predicted by the SC1, QP[0], and QP-NL theories are 0.73, 0.37, and 0.50 W/mK, respectively. Although no experimental $\kappa_{\mathrm{L}}$ value is available for $\alpha$-\cpb, 
we expect it would be similar to that of the orthorhombic phase $\kappa_{\mathrm{L}}\sim 0.4$ W/mK at 300 K~\cite{Lee_PNAS2017}. 
More recently, $\kappa_{\mathrm{L}}\sim 0.46$ W/mK at 500 K has been reported for another all-inorganic halide perovskite $\alpha$-CsSnBr$_3$~\cite{Xie_JACS2020}. Since the phonon frequencies of $\alpha$-\cpb and $\alpha$-CsSnBr$_3$ are quantitatively similar, a similar $\kappa_{\mathrm{L}}$ value is expected for $\alpha$-\cpb. \rr{Judging from these estimation, SC1 appears to overpredict} the LTC of $\alpha$-\cpb, and the predictions by the QP theories look more reasonable. 
\rr{We expect that calculations based on QP-NL are most reliable}, at least theoretically, because it best represents the peak positions of $A_q(\omega)$. This expectation should be validated by a future experimental study. 


To summarize, we developed a first-principles QP phonon theory that accounts for the bubble self-energy $\Sigma_q^{\mathrm{B}}[G,\Phi_3](\omega)$. The frequency-dependence of $\Sigma_q^{\mathrm{B}}[G,\Phi_3](\omega)$ was treated in three different ways: QP[0], QP[S], and QP-NL. By investingating the strongly anharmonic halide perovskite $\alpha$-\cpb, we demonstrated that these QP theories greatly improve the prediction accuracy of the cubic-to-tetragonal phase transition temperature and the \rr{phonon linewidth} as compared to the SC1 theory. We found that the QP-NL best represents the peak position of the spectral function $A_{q}(\omega)$. Therefore, the developed QP-NL theory offers an improved description of an \textit{effective} one-body Hamiltonian of strongly anharmonic systems and thereby paves the way to more accurate predictions of the structural phase transition temperature, \rr{phonon linewidth}, LTC, and electron-phonon coupling strength in various functional materials. 

\begin{acknowledgments}
This study is partly supported by JSPS KAKENHI Grant Number \rr{21K03424}.
\end{acknowledgments}


\begin{thebibliography}{60}%
    \makeatletter
    \providecommand \@ifxundefined [1]{%
     \@ifx{#1\undefined}
    }%
    \providecommand \@ifnum [1]{%
     \ifnum #1\expandafter \@firstoftwo
     \else \expandafter \@secondoftwo
     \fi
    }%
    \providecommand \@ifx [1]{%
     \ifx #1\expandafter \@firstoftwo
     \else \expandafter \@secondoftwo
     \fi
    }%
    \providecommand \natexlab [1]{#1}%
    \providecommand \enquote  [1]{``#1''}%
    \providecommand \bibnamefont  [1]{#1}%
    \providecommand \bibfnamefont [1]{#1}%
    \providecommand \citenamefont [1]{#1}%
    \providecommand \href@noop [0]{\@secondoftwo}%
    \providecommand \href [0]{\begingroup \@sanitize@url \@href}%
    \providecommand \@href[1]{\@@startlink{#1}\@@href}%
    \providecommand \@@href[1]{\endgroup#1\@@endlink}%
    \providecommand \@sanitize@url [0]{\catcode `\\12\catcode `\$12\catcode
      `\&12\catcode `\#12\catcode `\^12\catcode `\_12\catcode `\%12\relax}%
    \providecommand \@@startlink[1]{}%
    \providecommand \@@endlink[0]{}%
    \providecommand \url  [0]{\begingroup\@sanitize@url \@url }%
    \providecommand \@url [1]{\endgroup\@href {#1}{\urlprefix }}%
    \providecommand \urlprefix  [0]{URL }%
    \providecommand \Eprint [0]{\href }%
    \providecommand \doibase [0]{https://doi.org/}%
    \providecommand \selectlanguage [0]{\@gobble}%
    \providecommand \bibinfo  [0]{\@secondoftwo}%
    \providecommand \bibfield  [0]{\@secondoftwo}%
    \providecommand \translation [1]{[#1]}%
    \providecommand \BibitemOpen [0]{}%
    \providecommand \bibitemStop [0]{}%
    \providecommand \bibitemNoStop [0]{.\EOS\space}%
    \providecommand \EOS [0]{\spacefactor3000\relax}%
    \providecommand \BibitemShut  [1]{\csname bibitem#1\endcsname}%
    \let\auto@bib@innerbib\@empty
    \bibitem [{\citenamefont {Kozina}\ \emph {et~al.}(2019)\citenamefont {Kozina},
      \citenamefont {Fechner}, \citenamefont {Marsik}, \citenamefont {Driel},
      \citenamefont {Glownia}, \citenamefont {Bernhard}, \citenamefont {Radovic},
      \citenamefont {Zhu}, \citenamefont {Bonetti}, \citenamefont {Staub},\ and\
      \citenamefont {Hoffmann}}]{Kozina_NatPhys2019}%
      \BibitemOpen
      \bibfield  {author} {\bibinfo {author} {\bibfnamefont {M.}~\bibnamefont
      {Kozina}}, \bibinfo {author} {\bibfnamefont {M.}~\bibnamefont {Fechner}},
      \bibinfo {author} {\bibfnamefont {P.}~\bibnamefont {Marsik}}, \bibinfo
      {author} {\bibfnamefont {T.~v.}\ \bibnamefont {Driel}}, \bibinfo {author}
      {\bibfnamefont {J.~M.}\ \bibnamefont {Glownia}}, \bibinfo {author}
      {\bibfnamefont {C.}~\bibnamefont {Bernhard}}, \bibinfo {author}
      {\bibfnamefont {M.}~\bibnamefont {Radovic}}, \bibinfo {author} {\bibfnamefont
      {D.}~\bibnamefont {Zhu}}, \bibinfo {author} {\bibfnamefont {S.}~\bibnamefont
      {Bonetti}}, \bibinfo {author} {\bibfnamefont {U.}~\bibnamefont {Staub}},\
      and\ \bibinfo {author} {\bibfnamefont {M.~C.}\ \bibnamefont {Hoffmann}},\
      }\bibfield  {title} {\bibinfo {title} {{Terahertz-driven phonon upconversion
      in SrTiO$_3$}},\ }\href {https://doi.org/10.1038/s41567-018-0408-1}
      {\bibfield  {journal} {\bibinfo  {journal} {Nat. Phys.}\ }\textbf {\bibinfo
      {volume} {15}},\ \bibinfo {pages} {387} (\bibinfo {year} {2019})}\BibitemShut
      {NoStop}%
    \bibitem [{\citenamefont {Knoop}\ \emph {et~al.}(2020)\citenamefont {Knoop},
      \citenamefont {Purcell}, \citenamefont {Scheffler},\ and\ \citenamefont
      {Carbogno}}]{Knoop_PRM2020}%
      \BibitemOpen
      \bibfield  {author} {\bibinfo {author} {\bibfnamefont {F.}~\bibnamefont
      {Knoop}}, \bibinfo {author} {\bibfnamefont {T.~A.~R.}\ \bibnamefont
      {Purcell}}, \bibinfo {author} {\bibfnamefont {M.}~\bibnamefont {Scheffler}},\
      and\ \bibinfo {author} {\bibfnamefont {C.}~\bibnamefont {Carbogno}},\
      }\bibfield  {title} {\bibinfo {title} {{Anharmonicity measure for
      materials}},\ }\href {https://doi.org/10.1103/physrevmaterials.4.083809}
      {\bibfield  {journal} {\bibinfo  {journal} {Phys. Rev. Mater.}\ }\textbf
      {\bibinfo {volume} {4}},\ \bibinfo {pages} {083809} (\bibinfo {year}
      {2020})}\BibitemShut {NoStop}%
    \bibitem [{\citenamefont {Haque}\ \emph {et~al.}(2020)\citenamefont {Haque},
      \citenamefont {Kee}, \citenamefont {Villalva}, \citenamefont {Ong},\ and\
      \citenamefont {Baran}}]{Haque_AdvSci2020}%
      \BibitemOpen
      \bibfield  {author} {\bibinfo {author} {\bibfnamefont {M.~A.}\ \bibnamefont
      {Haque}}, \bibinfo {author} {\bibfnamefont {S.}~\bibnamefont {Kee}}, \bibinfo
      {author} {\bibfnamefont {D.~R.}\ \bibnamefont {Villalva}}, \bibinfo {author}
      {\bibfnamefont {W.}~\bibnamefont {Ong}},\ and\ \bibinfo {author}
      {\bibfnamefont {D.}~\bibnamefont {Baran}},\ }\bibfield  {title} {\bibinfo
      {title} {{Halide Perovskites: Thermal Transport and Prospects for
      Thermoelectricity}},\ }\href {https://doi.org/10.1002/advs.201903389}
      {\bibfield  {journal} {\bibinfo  {journal} {Adv. Sci.}\ }\textbf {\bibinfo
      {volume} {7}},\ \bibinfo {pages} {1903389} (\bibinfo {year}
      {2020})}\BibitemShut {NoStop}%
    \bibitem [{\citenamefont {Delaire}\ \emph {et~al.}(2011)\citenamefont
      {Delaire}, \citenamefont {Ma}, \citenamefont {Marty}, \citenamefont {May},
      \citenamefont {McGuire}, \citenamefont {Du}, \citenamefont {Singh},
      \citenamefont {Podlesnyak}, \citenamefont {Ehlers}, \citenamefont {Lumsden},\
      and\ \citenamefont {Sales}}]{Delaire_NatureMaterials2011}%
      \BibitemOpen
      \bibfield  {author} {\bibinfo {author} {\bibfnamefont {O.}~\bibnamefont
      {Delaire}}, \bibinfo {author} {\bibfnamefont {J.}~\bibnamefont {Ma}},
      \bibinfo {author} {\bibfnamefont {K.}~\bibnamefont {Marty}}, \bibinfo
      {author} {\bibfnamefont {A.~F.}\ \bibnamefont {May}}, \bibinfo {author}
      {\bibfnamefont {M.~A.}\ \bibnamefont {McGuire}}, \bibinfo {author}
      {\bibfnamefont {M.-H.}\ \bibnamefont {Du}}, \bibinfo {author} {\bibfnamefont
      {D.~J.}\ \bibnamefont {Singh}}, \bibinfo {author} {\bibfnamefont
      {A.}~\bibnamefont {Podlesnyak}}, \bibinfo {author} {\bibfnamefont
      {G.}~\bibnamefont {Ehlers}}, \bibinfo {author} {\bibfnamefont {M.~D.}\
      \bibnamefont {Lumsden}},\ and\ \bibinfo {author} {\bibfnamefont {B.~C.}\
      \bibnamefont {Sales}},\ }\bibfield  {title} {\bibinfo {title} {{Giant
      anharmonic phonon scattering in PbTe}},\ }\href
      {https://doi.org/10.1038/nmat3035} {\bibfield  {journal} {\bibinfo  {journal}
      {Nat. Mater.}\ }\textbf {\bibinfo {volume} {10}},\ \bibinfo {pages} {614}
      (\bibinfo {year} {2011})}\BibitemShut {NoStop}%
    \bibitem [{\citenamefont {Li}\ \emph {et~al.}(2015)\citenamefont {Li},
      \citenamefont {Hong}, \citenamefont {May}, \citenamefont {Bansal},
      \citenamefont {Chi}, \citenamefont {Hong}, \citenamefont {Ehlers},\ and\
      \citenamefont {Delaire}}]{Li_NaturePhysics2015}%
      \BibitemOpen
      \bibfield  {author} {\bibinfo {author} {\bibfnamefont {C.~W.}\ \bibnamefont
      {Li}}, \bibinfo {author} {\bibfnamefont {J.}~\bibnamefont {Hong}}, \bibinfo
      {author} {\bibfnamefont {A.~F.}\ \bibnamefont {May}}, \bibinfo {author}
      {\bibfnamefont {D.}~\bibnamefont {Bansal}}, \bibinfo {author} {\bibfnamefont
      {S.}~\bibnamefont {Chi}}, \bibinfo {author} {\bibfnamefont {T.}~\bibnamefont
      {Hong}}, \bibinfo {author} {\bibfnamefont {G.}~\bibnamefont {Ehlers}},\ and\
      \bibinfo {author} {\bibfnamefont {O.}~\bibnamefont {Delaire}},\ }\bibfield
      {title} {\bibinfo {title} {{Orbitally driven giant phonon anharmonicity in
      SnSe}},\ }\href {https://doi.org/10.1038/nphys3492} {\bibfield  {journal}
      {\bibinfo  {journal} {Nat. Phys.}\ }\textbf {\bibinfo {volume} {11}},\
      \bibinfo {pages} {1063} (\bibinfo {year} {2015})}\BibitemShut {NoStop}%
    \bibitem [{\citenamefont {Suekuni}\ \emph {et~al.}(2018)\citenamefont
      {Suekuni}, \citenamefont {Lee}, \citenamefont {Tanaka}, \citenamefont
      {Nishibori}, \citenamefont {Nakamura}, \citenamefont {Kasai}, \citenamefont
      {Mori}, \citenamefont {Usui}, \citenamefont {Ochi}, \citenamefont {Hasegawa},
      \citenamefont {Nakamura}, \citenamefont {Ohira‐Kawamura}, \citenamefont
      {Kikuchi}, \citenamefont {Kaneko}, \citenamefont {Nishiate}, \citenamefont
      {Hashikuni}, \citenamefont {Kosaka}, \citenamefont {Kuroki},\ and\
      \citenamefont {Takabatake}}]{Suekuni_AdvMater2018}%
      \BibitemOpen
      \bibfield  {author} {\bibinfo {author} {\bibfnamefont {K.}~\bibnamefont
      {Suekuni}}, \bibinfo {author} {\bibfnamefont {C.~H.}\ \bibnamefont {Lee}},
      \bibinfo {author} {\bibfnamefont {H.~I.}\ \bibnamefont {Tanaka}}, \bibinfo
      {author} {\bibfnamefont {E.}~\bibnamefont {Nishibori}}, \bibinfo {author}
      {\bibfnamefont {A.}~\bibnamefont {Nakamura}}, \bibinfo {author}
      {\bibfnamefont {H.}~\bibnamefont {Kasai}}, \bibinfo {author} {\bibfnamefont
      {H.}~\bibnamefont {Mori}}, \bibinfo {author} {\bibfnamefont {H.}~\bibnamefont
      {Usui}}, \bibinfo {author} {\bibfnamefont {M.}~\bibnamefont {Ochi}}, \bibinfo
      {author} {\bibfnamefont {T.}~\bibnamefont {Hasegawa}}, \bibinfo {author}
      {\bibfnamefont {M.}~\bibnamefont {Nakamura}}, \bibinfo {author}
      {\bibfnamefont {S.}~\bibnamefont {Ohira‐Kawamura}}, \bibinfo {author}
      {\bibfnamefont {T.}~\bibnamefont {Kikuchi}}, \bibinfo {author} {\bibfnamefont
      {K.}~\bibnamefont {Kaneko}}, \bibinfo {author} {\bibfnamefont
      {H.}~\bibnamefont {Nishiate}}, \bibinfo {author} {\bibfnamefont
      {K.}~\bibnamefont {Hashikuni}}, \bibinfo {author} {\bibfnamefont
      {Y.}~\bibnamefont {Kosaka}}, \bibinfo {author} {\bibfnamefont
      {K.}~\bibnamefont {Kuroki}},\ and\ \bibinfo {author} {\bibfnamefont
      {T.}~\bibnamefont {Takabatake}},\ }\bibfield  {title} {{\selectlanguage
      {English}\bibinfo {title} {{Retreat from Stress: Rattling in a Planar
      Coordination}}},\ }\href {https://doi.org/10.1002/adma.201706230} {\bibfield
      {journal} {\bibinfo  {journal} {Adv. Mater.}\ }\textbf {\bibinfo {volume}
      {30}},\ \bibinfo {pages} {1706230} (\bibinfo {year} {2018})}\BibitemShut
      {NoStop}%
    \bibitem [{\citenamefont {Errea}\ \emph {et~al.}(2015)\citenamefont {Errea},
      \citenamefont {Calandra}, \citenamefont {Pickard}, \citenamefont {Nelson},
      \citenamefont {Needs}, \citenamefont {Li}, \citenamefont {Liu}, \citenamefont
      {Zhang}, \citenamefont {Ma},\ and\ \citenamefont {Mauri}}]{Errea_PRL2015}%
      \BibitemOpen
      \bibfield  {author} {\bibinfo {author} {\bibfnamefont {I.}~\bibnamefont
      {Errea}}, \bibinfo {author} {\bibfnamefont {M.}~\bibnamefont {Calandra}},
      \bibinfo {author} {\bibfnamefont {C.~J.}\ \bibnamefont {Pickard}}, \bibinfo
      {author} {\bibfnamefont {J.}~\bibnamefont {Nelson}}, \bibinfo {author}
      {\bibfnamefont {R.~J.}\ \bibnamefont {Needs}}, \bibinfo {author}
      {\bibfnamefont {Y.}~\bibnamefont {Li}}, \bibinfo {author} {\bibfnamefont
      {H.}~\bibnamefont {Liu}}, \bibinfo {author} {\bibfnamefont {Y.}~\bibnamefont
      {Zhang}}, \bibinfo {author} {\bibfnamefont {Y.}~\bibnamefont {Ma}},\ and\
      \bibinfo {author} {\bibfnamefont {F.}~\bibnamefont {Mauri}},\ }\bibfield
      {title} {{\selectlanguage {English}\bibinfo {title} {{High-Pressure Hydrogen
      Sulfide from First Principles: A Strongly Anharmonic Phonon-Mediated
      Superconductor}}},\ }\href {https://doi.org/10.1103/physrevlett.114.157004}
      {\bibfield  {journal} {\bibinfo  {journal} {Phys. Rev. Lett.}\ }\textbf
      {\bibinfo {volume} {114}},\ \bibinfo {pages} {157004 } (\bibinfo {year}
      {2015})}\BibitemShut {NoStop}%
    \bibitem [{\citenamefont {Errea}\ \emph {et~al.}(2020)\citenamefont {Errea},
      \citenamefont {Belli}, \citenamefont {Monacelli}, \citenamefont {Sanna},
      \citenamefont {Koretsune}, \citenamefont {Tadano}, \citenamefont {Bianco},
      \citenamefont {Calandra}, \citenamefont {Arita}, \citenamefont {Mauri},\ and\
      \citenamefont {Flores-Livas}}]{Errea_Nature2020}%
      \BibitemOpen
      \bibfield  {author} {\bibinfo {author} {\bibfnamefont {I.}~\bibnamefont
      {Errea}}, \bibinfo {author} {\bibfnamefont {F.}~\bibnamefont {Belli}},
      \bibinfo {author} {\bibfnamefont {L.}~\bibnamefont {Monacelli}}, \bibinfo
      {author} {\bibfnamefont {A.}~\bibnamefont {Sanna}}, \bibinfo {author}
      {\bibfnamefont {T.}~\bibnamefont {Koretsune}}, \bibinfo {author}
      {\bibfnamefont {T.}~\bibnamefont {Tadano}}, \bibinfo {author} {\bibfnamefont
      {R.}~\bibnamefont {Bianco}}, \bibinfo {author} {\bibfnamefont
      {M.}~\bibnamefont {Calandra}}, \bibinfo {author} {\bibfnamefont
      {R.}~\bibnamefont {Arita}}, \bibinfo {author} {\bibfnamefont
      {F.}~\bibnamefont {Mauri}},\ and\ \bibinfo {author} {\bibfnamefont {J.~A.}\
      \bibnamefont {Flores-Livas}},\ }\bibfield  {title} {\bibinfo {title}
      {{Quantum crystal structure in the 250-kelvin superconducting lanthanum
      hydride}},\ }\href {https://doi.org/10.1038/s41586-020-1955-z} {\bibfield
      {journal} {\bibinfo  {journal} {Nature}\ }\textbf {\bibinfo {volume} {578}},\
      \bibinfo {pages} {66} (\bibinfo {year} {2020})}\BibitemShut {NoStop}%
    \bibitem [{\citenamefont {Huang}\ \emph {et~al.}(2017)\citenamefont {Huang},
      \citenamefont {Yuan}, \citenamefont {Shao},\ and\ \citenamefont
      {Yan}}]{Huang_NatReviewMater2017}%
      \BibitemOpen
      \bibfield  {author} {\bibinfo {author} {\bibfnamefont {J.}~\bibnamefont
      {Huang}}, \bibinfo {author} {\bibfnamefont {Y.}~\bibnamefont {Yuan}},
      \bibinfo {author} {\bibfnamefont {Y.}~\bibnamefont {Shao}},\ and\ \bibinfo
      {author} {\bibfnamefont {Y.}~\bibnamefont {Yan}},\ }\bibfield  {title}
      {\bibinfo {title} {{Understanding the physical properties of hybrid
      perovskites for photovoltaic applications}},\ }\href
      {https://doi.org/10.1038/natrevmats.2017.42} {\bibfield  {journal} {\bibinfo
      {journal} {Nat. Rev. Mater.}\ }\textbf {\bibinfo {volume} {2}},\ \bibinfo
      {pages} {17042} (\bibinfo {year} {2017})}\BibitemShut {NoStop}%
    \bibitem [{\citenamefont {Xie}\ \emph {et~al.}(2020)\citenamefont {Xie},
      \citenamefont {Hao}, \citenamefont {Bao}, \citenamefont {Slade},
      \citenamefont {Snyder}, \citenamefont {Wolverton},\ and\ \citenamefont
      {Kanatzidis}}]{Xie_JACS2020}%
      \BibitemOpen
      \bibfield  {author} {\bibinfo {author} {\bibfnamefont {H.}~\bibnamefont
      {Xie}}, \bibinfo {author} {\bibfnamefont {S.}~\bibnamefont {Hao}}, \bibinfo
      {author} {\bibfnamefont {J.}~\bibnamefont {Bao}}, \bibinfo {author}
      {\bibfnamefont {T.~J.}\ \bibnamefont {Slade}}, \bibinfo {author}
      {\bibfnamefont {G.~J.}\ \bibnamefont {Snyder}}, \bibinfo {author}
      {\bibfnamefont {C.}~\bibnamefont {Wolverton}},\ and\ \bibinfo {author}
      {\bibfnamefont {M.~G.}\ \bibnamefont {Kanatzidis}},\ }\bibfield  {title}
      {\bibinfo {title} {{All-Inorganic Halide Perovskites as Potential
      Thermoelectric Materials: Dynamic Cation off-Centering Induces Ultralow
      Thermal Conductivity}},\ }\href {https://doi.org/10.1021/jacs.0c03427}
      {\bibfield  {journal} {\bibinfo  {journal} {J. Am. Chem. Soc.}\ }\textbf
      {\bibinfo {volume} {142}},\ \bibinfo {pages} {9553} (\bibinfo {year}
      {2020})}\BibitemShut {NoStop}%
    \bibitem [{\citenamefont {Hellman}\ \emph {et~al.}(2011)\citenamefont
      {Hellman}, \citenamefont {Abrikosov},\ and\ \citenamefont
      {Simak}}]{Hellman_PRB2011}%
      \BibitemOpen
      \bibfield  {author} {\bibinfo {author} {\bibfnamefont {O.}~\bibnamefont
      {Hellman}}, \bibinfo {author} {\bibfnamefont {I.~A.}\ \bibnamefont
      {Abrikosov}},\ and\ \bibinfo {author} {\bibfnamefont {S.~I.}\ \bibnamefont
      {Simak}},\ }\bibfield  {title} {\bibinfo {title} {{Lattice dynamics of
      anharmonic solids from first principles}},\ }\href
      {https://doi.org/10.1103/physrevb.84.180301} {\bibfield  {journal} {\bibinfo
      {journal} {Phys. Rev. B}\ }\textbf {\bibinfo {volume} {84}},\ \bibinfo
      {pages} {180301} (\bibinfo {year} {2011})}\BibitemShut {NoStop}%
    \bibitem [{\citenamefont {Errea}\ \emph {et~al.}(2014)\citenamefont {Errea},
      \citenamefont {Calandra},\ and\ \citenamefont {Mauri}}]{Errea_PRB2014}%
      \BibitemOpen
      \bibfield  {author} {\bibinfo {author} {\bibfnamefont {I.}~\bibnamefont
      {Errea}}, \bibinfo {author} {\bibfnamefont {M.}~\bibnamefont {Calandra}},\
      and\ \bibinfo {author} {\bibfnamefont {F.}~\bibnamefont {Mauri}},\ }\bibfield
       {title} {{\selectlanguage {English}\bibinfo {title} {{Anharmonic free
      energies and phonon dispersions from the stochastic self-consistent harmonic
      approximation: Application to platinum and palladium hydrides}}},\ }\href
      {https://doi.org/10.1103/physrevb.89.064302} {\bibfield  {journal} {\bibinfo
      {journal} {Phys. Rev. B}\ }\textbf {\bibinfo {volume} {89}},\ \bibinfo
      {pages} {064302} (\bibinfo {year} {2014})}\BibitemShut {NoStop}%
    \bibitem [{\citenamefont {Tadano}\ and\ \citenamefont
      {Tsuneyuki}(2015)}]{Tadano_PRB2015}%
      \BibitemOpen
      \bibfield  {author} {\bibinfo {author} {\bibfnamefont {T.}~\bibnamefont
      {Tadano}}\ and\ \bibinfo {author} {\bibfnamefont {S.}~\bibnamefont
      {Tsuneyuki}},\ }\bibfield  {title} {{\selectlanguage {English}\bibinfo
      {title} {{Self-consistent phonon calculations of lattice dynamical properties
      in cubic SrTiO$_3$ with first-principles anharmonic force constants}}},\
      }\href {https://doi.org/10.1103/physrevb.92.054301} {\bibfield  {journal}
      {\bibinfo  {journal} {Phys. Rev. B}\ }\textbf {\bibinfo {volume} {92}},\
      \bibinfo {pages} {054301} (\bibinfo {year} {2015})}\BibitemShut {NoStop}%
    \bibitem [{\citenamefont {Roekeghem}\ \emph {et~al.}(2016)\citenamefont
      {Roekeghem}, \citenamefont {Carrete},\ and\ \citenamefont
      {Mingo}}]{Roekeghem_PRB2016}%
      \BibitemOpen
      \bibfield  {author} {\bibinfo {author} {\bibfnamefont {A.~v.}\ \bibnamefont
      {Roekeghem}}, \bibinfo {author} {\bibfnamefont {J.}~\bibnamefont {Carrete}},\
      and\ \bibinfo {author} {\bibfnamefont {N.}~\bibnamefont {Mingo}},\ }\bibfield
       {title} {{\selectlanguage {English}\bibinfo {title} {{Anomalous thermal
      conductivity and suppression of negative thermal expansion in ScF3}}},\
      }\href {https://doi.org/10.1103/physrevb.94.020303} {\bibfield  {journal}
      {\bibinfo  {journal} {Phys. Rev. B}\ }\textbf {\bibinfo {volume} {94}},\
      \bibinfo {pages} {020303 } (\bibinfo {year} {2016})}\BibitemShut {NoStop}%
    \bibitem [{\citenamefont {Ravichandran}\ and\ \citenamefont
      {Broido}(2018)}]{Ravichandran_PRB2018}%
      \BibitemOpen
      \bibfield  {author} {\bibinfo {author} {\bibfnamefont {N.~K.}\ \bibnamefont
      {Ravichandran}}\ and\ \bibinfo {author} {\bibfnamefont {D.}~\bibnamefont
      {Broido}},\ }\bibfield  {title} {{\selectlanguage {English}\bibinfo {title}
      {{Unified first-principles theory of thermal properties of insulators}}},\
      }\href {https://doi.org/10.1103/physrevb.98.085205} {\bibfield  {journal}
      {\bibinfo  {journal} {Phys. Rev. B}\ }\textbf {\bibinfo {volume} {98}},\
      \bibinfo {pages} {085205} (\bibinfo {year} {2018})}\BibitemShut {NoStop}%
    \bibitem [{\citenamefont {Klein}\ and\ \citenamefont
      {Horton}(1972)}]{Klein_JLTP1972}%
      \BibitemOpen
      \bibfield  {author} {\bibinfo {author} {\bibfnamefont {M.~L.}\ \bibnamefont
      {Klein}}\ and\ \bibinfo {author} {\bibfnamefont {G.~K.}\ \bibnamefont
      {Horton}},\ }\bibfield  {title} {\bibinfo {title} {{The rise of
      self-consistent phonon theory}},\ }\href {https://doi.org/10.1007/bf00654839}
      {\bibfield  {journal} {\bibinfo  {journal} {J. Low Temp. Phys.}\ }\textbf
      {\bibinfo {volume} {9}},\ \bibinfo {pages} {151} (\bibinfo {year}
      {1972})}\BibitemShut {NoStop}%
    \bibitem [{\citenamefont {Tadano}\ and\ \citenamefont
      {Tsuneyuki}(2018{\natexlab{a}})}]{Tadano_JPSJ2018}%
      \BibitemOpen
      \bibfield  {author} {\bibinfo {author} {\bibfnamefont {T.}~\bibnamefont
      {Tadano}}\ and\ \bibinfo {author} {\bibfnamefont {S.}~\bibnamefont
      {Tsuneyuki}},\ }\bibfield  {title} {{\selectlanguage {English}\bibinfo
      {title} {{First-Principles Lattice Dynamics Method for Strongly Anharmonic
      Crystals}}},\ }\href {https://doi.org/10.7566/jpsj.87.041015} {\bibfield
      {journal} {\bibinfo  {journal} {J. Phys. Soc. Jpn.}\ }\textbf {\bibinfo
      {volume} {87}},\ \bibinfo {pages} {041015 } (\bibinfo {year}
      {2018}{\natexlab{a}})}\BibitemShut {NoStop}%
    \bibitem [{\citenamefont {Kang}\ \emph {et~al.}(2019)\citenamefont {Kang},
      \citenamefont {Wu}, \citenamefont {Li},\ and\ \citenamefont
      {Hu}}]{Kang_NanoLett2019}%
      \BibitemOpen
      \bibfield  {author} {\bibinfo {author} {\bibfnamefont {J.~S.}\ \bibnamefont
      {Kang}}, \bibinfo {author} {\bibfnamefont {H.}~\bibnamefont {Wu}}, \bibinfo
      {author} {\bibfnamefont {M.}~\bibnamefont {Li}},\ and\ \bibinfo {author}
      {\bibfnamefont {Y.}~\bibnamefont {Hu}},\ }\bibfield  {title} {\bibinfo
      {title} {{Intrinsic Low Thermal Conductivity and Phonon Renormalization Due
      to Strong Anharmonicity of Single-Crystal Tin Selenide}},\ }\href
      {https://doi.org/10.1021/acs.nanolett.9b01056} {\bibfield  {journal}
      {\bibinfo  {journal} {Nano Lett.}\ }\textbf {\bibinfo {volume} {19}},\
      \bibinfo {pages} {4941} (\bibinfo {year} {2019})}\BibitemShut {NoStop}%
    \bibitem [{\citenamefont {Xia}\ \emph {et~al.}(2020{\natexlab{a}})\citenamefont
      {Xia}, \citenamefont {Pal}, \citenamefont {He}, \citenamefont {Ozoliņš},\
      and\ \citenamefont {Wolverton}}]{Xia_PRL2020}%
      \BibitemOpen
      \bibfield  {author} {\bibinfo {author} {\bibfnamefont {Y.}~\bibnamefont
      {Xia}}, \bibinfo {author} {\bibfnamefont {K.}~\bibnamefont {Pal}}, \bibinfo
      {author} {\bibfnamefont {J.}~\bibnamefont {He}}, \bibinfo {author}
      {\bibfnamefont {V.}~\bibnamefont {Ozoliņš}},\ and\ \bibinfo {author}
      {\bibfnamefont {C.}~\bibnamefont {Wolverton}},\ }\bibfield  {title} {\bibinfo
      {title} {{Particlelike Phonon Propagation Dominates Ultralow Lattice Thermal
      Conductivity in Crystalline Tl$_3$VSe$_4$}},\ }\href
      {https://doi.org/10.1103/physrevlett.124.065901} {\bibfield  {journal}
      {\bibinfo  {journal} {Phys. Rev. Lett.}\ }\textbf {\bibinfo {volume} {124}},\
      \bibinfo {pages} {065901} (\bibinfo {year} {2020}{\natexlab{a}})}\BibitemShut
      {NoStop}%
    \bibitem [{\citenamefont {Xia}\ \emph {et~al.}(2020{\natexlab{b}})\citenamefont
      {Xia}, \citenamefont {Hegde}, \citenamefont {Pal}, \citenamefont {Hua},
      \citenamefont {Gaines}, \citenamefont {Patel}, \citenamefont {He},
      \citenamefont {Aykol},\ and\ \citenamefont {Wolverton}}]{Xia_PRX2020}%
      \BibitemOpen
      \bibfield  {author} {\bibinfo {author} {\bibfnamefont {Y.}~\bibnamefont
      {Xia}}, \bibinfo {author} {\bibfnamefont {V.~I.}\ \bibnamefont {Hegde}},
      \bibinfo {author} {\bibfnamefont {K.}~\bibnamefont {Pal}}, \bibinfo {author}
      {\bibfnamefont {X.}~\bibnamefont {Hua}}, \bibinfo {author} {\bibfnamefont
      {D.}~\bibnamefont {Gaines}}, \bibinfo {author} {\bibfnamefont
      {S.}~\bibnamefont {Patel}}, \bibinfo {author} {\bibfnamefont
      {J.}~\bibnamefont {He}}, \bibinfo {author} {\bibfnamefont {M.}~\bibnamefont
      {Aykol}},\ and\ \bibinfo {author} {\bibfnamefont {C.}~\bibnamefont
      {Wolverton}},\ }\bibfield  {title} {\bibinfo {title} {{High-Throughput Study
      of Lattice Thermal Conductivity in Binary Rocksalt and Zinc Blende Compounds
      Including Higher-Order Anharmonicity}},\ }\href
      {https://doi.org/10.1103/physrevx.10.041029} {\bibfield  {journal} {\bibinfo
      {journal} {Phys. Rev. X}\ }\textbf {\bibinfo {volume} {10}},\ \bibinfo
      {pages} {041029} (\bibinfo {year} {2020}{\natexlab{b}})}\BibitemShut
      {NoStop}%
    \bibitem [{\citenamefont {Kawano}\ \emph {et~al.}(2021)\citenamefont {Kawano},
      \citenamefont {Tadano},\ and\ \citenamefont {Iikubo}}]{Kawano_JPCC2021}%
      \BibitemOpen
      \bibfield  {author} {\bibinfo {author} {\bibfnamefont {S.}~\bibnamefont
      {Kawano}}, \bibinfo {author} {\bibfnamefont {T.}~\bibnamefont {Tadano}},\
      and\ \bibinfo {author} {\bibfnamefont {S.}~\bibnamefont {Iikubo}},\
      }\bibfield  {title} {\bibinfo {title} {{Effect of Halogen Ions on the Low
      Thermal Conductivity of Cesium Halide Perovskite}},\ }\href
      {https://doi.org/10.1021/acs.jpcc.0c08324} {\bibfield  {journal} {\bibinfo
      {journal} {J. Phys. Chem. C}\ }\textbf {\bibinfo {volume} {125}},\ \bibinfo
      {pages} {91} (\bibinfo {year} {2021})}\BibitemShut {NoStop}%
    \bibitem [{\citenamefont {Zhao}\ \emph {et~al.}(2020)\citenamefont {Zhao},
      \citenamefont {Lian}, \citenamefont {Zeng}, \citenamefont {Dai},
      \citenamefont {Meng},\ and\ \citenamefont {Ni}}]{Zhao_PRB2020}%
      \BibitemOpen
      \bibfield  {author} {\bibinfo {author} {\bibfnamefont {Y.}~\bibnamefont
      {Zhao}}, \bibinfo {author} {\bibfnamefont {C.}~\bibnamefont {Lian}}, \bibinfo
      {author} {\bibfnamefont {S.}~\bibnamefont {Zeng}}, \bibinfo {author}
      {\bibfnamefont {Z.}~\bibnamefont {Dai}}, \bibinfo {author} {\bibfnamefont
      {S.}~\bibnamefont {Meng}},\ and\ \bibinfo {author} {\bibfnamefont
      {J.}~\bibnamefont {Ni}},\ }\bibfield  {title} {\bibinfo {title} {{Anomalous
      electronic and thermoelectric transport properties in cubic Rb$_3$AuO
      antiperovskite}},\ }\href {https://doi.org/10.1103/physrevb.102.094314}
      {\bibfield  {journal} {\bibinfo  {journal} {Phys. Rev. B}\ }\textbf {\bibinfo
      {volume} {102}},\ \bibinfo {pages} {094314} (\bibinfo {year}
      {2020})}\BibitemShut {NoStop}%
    \bibitem [{\citenamefont {Patrick}\ \emph {et~al.}(2015)\citenamefont
      {Patrick}, \citenamefont {Jacobsen},\ and\ \citenamefont
      {Thygesen}}]{Patrick_PRB2015}%
      \BibitemOpen
      \bibfield  {author} {\bibinfo {author} {\bibfnamefont {C.~E.}\ \bibnamefont
      {Patrick}}, \bibinfo {author} {\bibfnamefont {K.~W.}\ \bibnamefont
      {Jacobsen}},\ and\ \bibinfo {author} {\bibfnamefont {K.~S.}\ \bibnamefont
      {Thygesen}},\ }\bibfield  {title} {{\selectlanguage {English}\bibinfo {title}
      {{Anharmonic stabilization and band gap renormalization in the perovskite
      CsSnI$_3$}}},\ }\href {https://doi.org/10.1103/physrevb.92.201205} {\bibfield
       {journal} {\bibinfo  {journal} {Phys. Rev. B}\ }\textbf {\bibinfo {volume}
      {92}},\ \bibinfo {pages} {201205 } (\bibinfo {year} {2015})}\BibitemShut
      {NoStop}%
    \bibitem [{\citenamefont {Wu}\ \emph {et~al.}(2020)\citenamefont {Wu},
      \citenamefont {Saidi}, \citenamefont {Wuenschell}, \citenamefont {Tadano},
      \citenamefont {Ohodnicki}, \citenamefont {Chorpening},\ and\ \citenamefont
      {Duan}}]{Wu_JPCL2020}%
      \BibitemOpen
      \bibfield  {author} {\bibinfo {author} {\bibfnamefont {Y.-N.}\ \bibnamefont
      {Wu}}, \bibinfo {author} {\bibfnamefont {W.~A.}\ \bibnamefont {Saidi}},
      \bibinfo {author} {\bibfnamefont {J.~K.}\ \bibnamefont {Wuenschell}},
      \bibinfo {author} {\bibfnamefont {T.}~\bibnamefont {Tadano}}, \bibinfo
      {author} {\bibfnamefont {P.}~\bibnamefont {Ohodnicki}}, \bibinfo {author}
      {\bibfnamefont {B.}~\bibnamefont {Chorpening}},\ and\ \bibinfo {author}
      {\bibfnamefont {Y.}~\bibnamefont {Duan}},\ }\bibfield  {title} {\bibinfo
      {title} {{Anharmonicity Explains Temperature Renormalization Effects of the
      Band Gap in SrTiO$_3$}},\ }\href
      {https://doi.org/10.1021/acs.jpclett.0c00183} {\bibfield  {journal} {\bibinfo
       {journal} {J. Phys. Chem. Lett.}\ }\textbf {\bibinfo {volume} {11}},\
      \bibinfo {pages} {2518} (\bibinfo {year} {2020})}\BibitemShut {NoStop}%
    \bibitem [{\citenamefont {Oba}\ \emph {et~al.}(2019)\citenamefont {Oba},
      \citenamefont {Tadano}, \citenamefont {Akashi},\ and\ \citenamefont
      {Tsuneyuki}}]{Oba_PRM2019}%
      \BibitemOpen
      \bibfield  {author} {\bibinfo {author} {\bibfnamefont {Y.}~\bibnamefont
      {Oba}}, \bibinfo {author} {\bibfnamefont {T.}~\bibnamefont {Tadano}},
      \bibinfo {author} {\bibfnamefont {R.}~\bibnamefont {Akashi}},\ and\ \bibinfo
      {author} {\bibfnamefont {S.}~\bibnamefont {Tsuneyuki}},\ }\bibfield  {title}
      {{\selectlanguage {English}\bibinfo {title} {{First-principles study of
      phonon anharmonicity and negative thermal expansion in ScF$_3$}}},\ }\href
      {https://doi.org/10.1103/physrevmaterials.3.033601} {\bibfield  {journal}
      {\bibinfo  {journal} {Phys. Rev. Mater.}\ }\textbf {\bibinfo {volume} {3}},\
      \bibinfo {pages} {033601} (\bibinfo {year} {2019})}\BibitemShut {NoStop}%
    \bibitem [{\citenamefont {Kwon}\ \emph {et~al.}(2020)\citenamefont {Kwon},
      \citenamefont {Xia}, \citenamefont {Zhou},\ and\ \citenamefont
      {Han}}]{Kwon_PRB2020}%
      \BibitemOpen
      \bibfield  {author} {\bibinfo {author} {\bibfnamefont {C.}~\bibnamefont
      {Kwon}}, \bibinfo {author} {\bibfnamefont {Y.}~\bibnamefont {Xia}}, \bibinfo
      {author} {\bibfnamefont {F.}~\bibnamefont {Zhou}},\ and\ \bibinfo {author}
      {\bibfnamefont {B.}~\bibnamefont {Han}},\ }\bibfield  {title} {\bibinfo
      {title} {{Dominant effect of anharmonicity on the equation of state and
      thermal conductivity of MgO under extreme conditions}},\ }\href
      {https://doi.org/10.1103/physrevb.102.184309} {\bibfield  {journal} {\bibinfo
       {journal} {Phys. Rev. B}\ }\textbf {\bibinfo {volume} {102}},\ \bibinfo
      {pages} {184309} (\bibinfo {year} {2020})}\BibitemShut {NoStop}%
    \bibitem [{\citenamefont {Sano}\ \emph {et~al.}(2016)\citenamefont {Sano},
      \citenamefont {Koretsune}, \citenamefont {Tadano}, \citenamefont {Akashi},\
      and\ \citenamefont {Arita}}]{Sano_PRB2016}%
      \BibitemOpen
      \bibfield  {author} {\bibinfo {author} {\bibfnamefont {W.}~\bibnamefont
      {Sano}}, \bibinfo {author} {\bibfnamefont {T.}~\bibnamefont {Koretsune}},
      \bibinfo {author} {\bibfnamefont {T.}~\bibnamefont {Tadano}}, \bibinfo
      {author} {\bibfnamefont {R.}~\bibnamefont {Akashi}},\ and\ \bibinfo {author}
      {\bibfnamefont {R.}~\bibnamefont {Arita}},\ }\bibfield  {title} {\bibinfo
      {title} {{Effect of Van Hove singularities on high-Tc superconductivity in
      H$_3$S}},\ }\href {https://doi.org/10.1103/physrevb.93.094525} {\bibfield
      {journal} {\bibinfo  {journal} {Phys. Rev. B}\ }\textbf {\bibinfo {volume}
      {93}},\ \bibinfo {pages} {094525} (\bibinfo {year} {2016})}\BibitemShut
      {NoStop}%
    \bibitem [{\citenamefont {Martin}\ \emph {et~al.}(2016)\citenamefont {Martin},
      \citenamefont {Reining},\ and\ \citenamefont
      {Ceperley}}]{martin2016interacting}%
      \BibitemOpen
      \bibfield  {author} {\bibinfo {author} {\bibfnamefont {R.}~\bibnamefont
      {Martin}}, \bibinfo {author} {\bibfnamefont {L.}~\bibnamefont {Reining}},\
      and\ \bibinfo {author} {\bibfnamefont {D.}~\bibnamefont {Ceperley}},\ }\href
      {https://books.google.co.jp/books?id=UAapDAAAQBAJ} {\emph {\bibinfo {title}
      {Interacting Electrons: Theory and Computational Approaches}}}\ (\bibinfo
      {publisher} {Cambridge University Press},\ \bibinfo {year}
      {2016})\BibitemShut {NoStop}%
    \bibitem [{\citenamefont {Hirotsu}\ \emph {et~al.}(1974)\citenamefont
      {Hirotsu}, \citenamefont {Harada}, \citenamefont {Iizumi},\ and\
      \citenamefont {Gesi}}]{Hirotsu_JPSJ1974}%
      \BibitemOpen
      \bibfield  {author} {\bibinfo {author} {\bibfnamefont {S.}~\bibnamefont
      {Hirotsu}}, \bibinfo {author} {\bibfnamefont {J.}~\bibnamefont {Harada}},
      \bibinfo {author} {\bibfnamefont {M.}~\bibnamefont {Iizumi}},\ and\ \bibinfo
      {author} {\bibfnamefont {K.}~\bibnamefont {Gesi}},\ }\bibfield  {title}
      {{\selectlanguage {English}\bibinfo {title} {{Structural Phase Transitions in
      CsPbBr$_3$}}},\ }\href {https://doi.org/10.1143/jpsj.37.1393} {\bibfield
      {journal} {\bibinfo  {journal} {J. Phys. Soc. Jpn.}\ }\textbf {\bibinfo
      {volume} {37}},\ \bibinfo {pages} {1393 } (\bibinfo {year}
      {1974})}\BibitemShut {NoStop}%
    \bibitem [{\citenamefont {Simoncelli}\ \emph {et~al.}(2019)\citenamefont
      {Simoncelli}, \citenamefont {Marzari},\ and\ \citenamefont
      {Mauri}}]{Simoncelli_NatPhys2019}%
      \BibitemOpen
      \bibfield  {author} {\bibinfo {author} {\bibfnamefont {M.}~\bibnamefont
      {Simoncelli}}, \bibinfo {author} {\bibfnamefont {N.}~\bibnamefont
      {Marzari}},\ and\ \bibinfo {author} {\bibfnamefont {F.}~\bibnamefont
      {Mauri}},\ }\bibfield  {title} {{\selectlanguage {English}\bibinfo {title}
      {{Unified theory of thermal transport in crystals and glasses}}},\ }\href
      {https://doi.org/10.1038/s41567-019-0520-x} {\bibfield  {journal} {\bibinfo
      {journal} {Nat. Phys.}\ }\textbf {\bibinfo {volume} {15}},\ \bibinfo {pages}
      {809} (\bibinfo {year} {2019})}\BibitemShut {NoStop}%
    \bibitem [{\citenamefont {Aseginolaza}\ \emph {et~al.}(2019)\citenamefont
      {Aseginolaza}, \citenamefont {Bianco}, \citenamefont {Monacelli},
      \citenamefont {Paulatto}, \citenamefont {Calandra}, \citenamefont {Mauri},
      \citenamefont {Bergara},\ and\ \citenamefont {Errea}}]{Aseginolaza_PRL2019}%
      \BibitemOpen
      \bibfield  {author} {\bibinfo {author} {\bibfnamefont {U.}~\bibnamefont
      {Aseginolaza}}, \bibinfo {author} {\bibfnamefont {R.}~\bibnamefont {Bianco}},
      \bibinfo {author} {\bibfnamefont {L.}~\bibnamefont {Monacelli}}, \bibinfo
      {author} {\bibfnamefont {L.}~\bibnamefont {Paulatto}}, \bibinfo {author}
      {\bibfnamefont {M.}~\bibnamefont {Calandra}}, \bibinfo {author}
      {\bibfnamefont {F.}~\bibnamefont {Mauri}}, \bibinfo {author} {\bibfnamefont
      {A.}~\bibnamefont {Bergara}},\ and\ \bibinfo {author} {\bibfnamefont
      {I.}~\bibnamefont {Errea}},\ }\bibfield  {title} {{\selectlanguage
      {English}\bibinfo {title} {{Phonon Collapse and Second-Order Phase Transition
      in Thermoelectric SnSe}}},\ }\href
      {https://doi.org/10.1103/physrevlett.122.075901} {\bibfield  {journal}
      {\bibinfo  {journal} {Phys. Rev. Lett.}\ }\textbf {\bibinfo {volume} {122}},\
      \bibinfo {pages} {075901} (\bibinfo {year} {2019})}\BibitemShut {NoStop}%
    \bibitem [{Not()}]{Note1}%
      \BibitemOpen
      \href@noop {} {}\bibinfo {note} {Here we neglect the off-diagonal components
      of the bubble self-energy, which are less significant than the diagonal
      terms. Extension of including the off-diagonal components is
      possible.}\BibitemShut {Stop}%
    \bibitem [{\citenamefont {Bianco}\ \emph {et~al.}(2017)\citenamefont {Bianco},
      \citenamefont {Errea}, \citenamefont {Paulatto}, \citenamefont {Calandra},\
      and\ \citenamefont {Mauri}}]{Bianco_PRB2017}%
      \BibitemOpen
      \bibfield  {author} {\bibinfo {author} {\bibfnamefont {R.}~\bibnamefont
      {Bianco}}, \bibinfo {author} {\bibfnamefont {I.}~\bibnamefont {Errea}},
      \bibinfo {author} {\bibfnamefont {L.}~\bibnamefont {Paulatto}}, \bibinfo
      {author} {\bibfnamefont {M.}~\bibnamefont {Calandra}},\ and\ \bibinfo
      {author} {\bibfnamefont {F.}~\bibnamefont {Mauri}},\ }\bibfield  {title}
      {{\selectlanguage {English}\bibinfo {title} {{Second-order structural phase
      transitions, free energy curvature, and temperature-dependent anharmonic
      phonons in the self-consistent harmonic approximation: Theory and stochastic
      implementation}}},\ }\href {https://doi.org/10.1103/physrevb.96.014111}
      {\bibfield  {journal} {\bibinfo  {journal} {Phys. Rev. B}\ }\textbf {\bibinfo
      {volume} {96}},\ \bibinfo {pages} {014111} (\bibinfo {year}
      {2017})}\BibitemShut {NoStop}%
    \bibitem [{\citenamefont {Schilfgaarde}\ \emph {et~al.}(2006)\citenamefont
      {Schilfgaarde}, \citenamefont {Kotani},\ and\ \citenamefont
      {Faleev}}]{Schilfgaarde_PRL2006}%
      \BibitemOpen
      \bibfield  {author} {\bibinfo {author} {\bibfnamefont {M.~v.}\ \bibnamefont
      {Schilfgaarde}}, \bibinfo {author} {\bibfnamefont {T.}~\bibnamefont
      {Kotani}},\ and\ \bibinfo {author} {\bibfnamefont {S.}~\bibnamefont
      {Faleev}},\ }\bibfield  {title} {{\selectlanguage {English}\bibinfo {title}
      {{Quasiparticle Self-Consistent GW Theory}}},\ }\href
      {https://doi.org/10.1103/physrevlett.96.226402} {\bibfield  {journal}
      {\bibinfo  {journal} {Phys. Rev. Lett.}\ }\textbf {\bibinfo {volume} {96}},\
      \bibinfo {pages} {226402} (\bibinfo {year} {2006})}\BibitemShut {NoStop}%
    \bibitem [{\citenamefont {López}\ \emph {et~al.}(2020)\citenamefont {López},
      \citenamefont {Abia}, \citenamefont {Alvarez-Galván}, \citenamefont {Hong},
      \citenamefont {Mart\'{i}nez-Huerta}, \citenamefont {Serrano-Sánchez},
      \citenamefont {Carrascoso}, \citenamefont {Castellanos-Gómez}, \citenamefont
      {Fernández-D\'{i}az},\ and\ \citenamefont {Alonso}}]{Lopez_Omega2020}%
      \BibitemOpen
      \bibfield  {author} {\bibinfo {author} {\bibfnamefont {C.~A.}\ \bibnamefont
      {López}}, \bibinfo {author} {\bibfnamefont {C.}~\bibnamefont {Abia}},
      \bibinfo {author} {\bibfnamefont {M.~C.}\ \bibnamefont {Alvarez-Galván}},
      \bibinfo {author} {\bibfnamefont {B.-K.}\ \bibnamefont {Hong}}, \bibinfo
      {author} {\bibfnamefont {M.~V.}\ \bibnamefont {Mart\'{i}nez-Huerta}},
      \bibinfo {author} {\bibfnamefont {F.}~\bibnamefont {Serrano-Sánchez}},
      \bibinfo {author} {\bibfnamefont {F.}~\bibnamefont {Carrascoso}}, \bibinfo
      {author} {\bibfnamefont {A.}~\bibnamefont {Castellanos-Gómez}}, \bibinfo
      {author} {\bibfnamefont {M.~T.}\ \bibnamefont {Fernández-D\'{i}az}},\ and\
      \bibinfo {author} {\bibfnamefont {J.~A.}\ \bibnamefont {Alonso}},\ }\bibfield
       {title} {\bibinfo {title} {{Crystal Structure Features of CsPbBr$_3$
      Perovskite Prepared by Mechanochemical Synthesis}},\ }\href
      {https://doi.org/10.1021/acsomega.9b04248} {\bibfield  {journal} {\bibinfo
      {journal} {ACS Omega}\ }\textbf {\bibinfo {volume} {5}},\ \bibinfo {pages}
      {5931} (\bibinfo {year} {2020})}\BibitemShut {NoStop}%
    \bibitem [{\citenamefont {Giannozzi}\ \emph {et~al.}(2017)\citenamefont
      {Giannozzi}, \citenamefont {Andreussi}, \citenamefont {Brumme}, \citenamefont
      {Bunau}, \citenamefont {Nardelli}, \citenamefont {Calandra}, \citenamefont
      {Car}, \citenamefont {Cavazzoni}, \citenamefont {Ceresoli}, \citenamefont
      {Cococcioni}, \citenamefont {Colonna}, \citenamefont {Carnimeo},
      \citenamefont {Corso}, \citenamefont {Gironcoli}, \citenamefont {Delugas},
      \citenamefont {Jr}, \citenamefont {Ferretti}, \citenamefont {Floris},
      \citenamefont {Fratesi}, \citenamefont {Fugallo}, \citenamefont {Gebauer},
      \citenamefont {Gerstmann}, \citenamefont {Giustino}, \citenamefont {Gorni},
      \citenamefont {Jia}, \citenamefont {Kawamura}, \citenamefont {Ko},
      \citenamefont {Kokalj}, \citenamefont {Küçükbenli}, \citenamefont
      {Lazzeri}, \citenamefont {Marsili}, \citenamefont {Marzari}, \citenamefont
      {Mauri}, \citenamefont {Nguyen}, \citenamefont {Nguyen}, \citenamefont
      {Otero-de-la Roza}, \citenamefont {Paulatto}, \citenamefont {Poncé},
      \citenamefont {Rocca}, \citenamefont {Sabatini}, \citenamefont {Santra},
      \citenamefont {Schlipf}, \citenamefont {Seitsonen}, \citenamefont {Smogunov},
      \citenamefont {Timrov}, \citenamefont {Thonhauser}, \citenamefont {Umari},
      \citenamefont {Vast}, \citenamefont {Wu},\ and\ \citenamefont
      {Baroni}}]{QE2017}%
      \BibitemOpen
      \bibfield  {author} {\bibinfo {author} {\bibfnamefont {P.}~\bibnamefont
      {Giannozzi}}, \bibinfo {author} {\bibfnamefont {O.}~\bibnamefont
      {Andreussi}}, \bibinfo {author} {\bibfnamefont {T.}~\bibnamefont {Brumme}},
      \bibinfo {author} {\bibfnamefont {O.}~\bibnamefont {Bunau}}, \bibinfo
      {author} {\bibfnamefont {M.~B.}\ \bibnamefont {Nardelli}}, \bibinfo {author}
      {\bibfnamefont {M.}~\bibnamefont {Calandra}}, \bibinfo {author}
      {\bibfnamefont {R.}~\bibnamefont {Car}}, \bibinfo {author} {\bibfnamefont
      {C.}~\bibnamefont {Cavazzoni}}, \bibinfo {author} {\bibfnamefont
      {D.}~\bibnamefont {Ceresoli}}, \bibinfo {author} {\bibfnamefont
      {M.}~\bibnamefont {Cococcioni}}, \bibinfo {author} {\bibfnamefont
      {N.}~\bibnamefont {Colonna}}, \bibinfo {author} {\bibfnamefont
      {I.}~\bibnamefont {Carnimeo}}, \bibinfo {author} {\bibfnamefont {A.~D.}\
      \bibnamefont {Corso}}, \bibinfo {author} {\bibfnamefont {S.~d.}\ \bibnamefont
      {Gironcoli}}, \bibinfo {author} {\bibfnamefont {P.}~\bibnamefont {Delugas}},
      \bibinfo {author} {\bibfnamefont {R.~A.~D.}\ \bibnamefont {Jr}}, \bibinfo
      {author} {\bibfnamefont {A.}~\bibnamefont {Ferretti}}, \bibinfo {author}
      {\bibfnamefont {A.}~\bibnamefont {Floris}}, \bibinfo {author} {\bibfnamefont
      {G.}~\bibnamefont {Fratesi}}, \bibinfo {author} {\bibfnamefont
      {G.}~\bibnamefont {Fugallo}}, \bibinfo {author} {\bibfnamefont
      {R.}~\bibnamefont {Gebauer}}, \bibinfo {author} {\bibfnamefont
      {U.}~\bibnamefont {Gerstmann}}, \bibinfo {author} {\bibfnamefont
      {F.}~\bibnamefont {Giustino}}, \bibinfo {author} {\bibfnamefont
      {T.}~\bibnamefont {Gorni}}, \bibinfo {author} {\bibfnamefont
      {J.}~\bibnamefont {Jia}}, \bibinfo {author} {\bibfnamefont {M.}~\bibnamefont
      {Kawamura}}, \bibinfo {author} {\bibfnamefont {H.-Y.}\ \bibnamefont {Ko}},
      \bibinfo {author} {\bibfnamefont {A.}~\bibnamefont {Kokalj}}, \bibinfo
      {author} {\bibfnamefont {E.}~\bibnamefont {Küçükbenli}}, \bibinfo {author}
      {\bibfnamefont {M.}~\bibnamefont {Lazzeri}}, \bibinfo {author} {\bibfnamefont
      {M.}~\bibnamefont {Marsili}}, \bibinfo {author} {\bibfnamefont
      {N.}~\bibnamefont {Marzari}}, \bibinfo {author} {\bibfnamefont
      {F.}~\bibnamefont {Mauri}}, \bibinfo {author} {\bibfnamefont {N.~L.}\
      \bibnamefont {Nguyen}}, \bibinfo {author} {\bibfnamefont {H.-V.}\
      \bibnamefont {Nguyen}}, \bibinfo {author} {\bibfnamefont {A.}~\bibnamefont
      {Otero-de-la Roza}}, \bibinfo {author} {\bibfnamefont {L.}~\bibnamefont
      {Paulatto}}, \bibinfo {author} {\bibfnamefont {S.}~\bibnamefont {Poncé}},
      \bibinfo {author} {\bibfnamefont {D.}~\bibnamefont {Rocca}}, \bibinfo
      {author} {\bibfnamefont {R.}~\bibnamefont {Sabatini}}, \bibinfo {author}
      {\bibfnamefont {B.}~\bibnamefont {Santra}}, \bibinfo {author} {\bibfnamefont
      {M.}~\bibnamefont {Schlipf}}, \bibinfo {author} {\bibfnamefont {A.~P.}\
      \bibnamefont {Seitsonen}}, \bibinfo {author} {\bibfnamefont {A.}~\bibnamefont
      {Smogunov}}, \bibinfo {author} {\bibfnamefont {I.}~\bibnamefont {Timrov}},
      \bibinfo {author} {\bibfnamefont {T.}~\bibnamefont {Thonhauser}}, \bibinfo
      {author} {\bibfnamefont {P.}~\bibnamefont {Umari}}, \bibinfo {author}
      {\bibfnamefont {N.}~\bibnamefont {Vast}}, \bibinfo {author} {\bibfnamefont
      {X.}~\bibnamefont {Wu}},\ and\ \bibinfo {author} {\bibfnamefont
      {S.}~\bibnamefont {Baroni}},\ }\bibfield  {title} {\bibinfo {title}
      {{Advanced capabilities for materials modelling with Quantum ESPRESSO}},\
      }\href {https://doi.org/10.1088/1361-648x/aa8f79} {\bibfield  {journal}
      {\bibinfo  {journal} {J. Phys.: Condens. Matter}\ }\textbf {\bibinfo {volume}
      {29}},\ \bibinfo {pages} {465901} (\bibinfo {year} {2017})}\BibitemShut
      {NoStop}%
    \bibitem [{\citenamefont {Perdew}\ \emph {et~al.}(2008)\citenamefont {Perdew},
      \citenamefont {Ruzsinszky}, \citenamefont {Csonka}, \citenamefont {Vydrov},
      \citenamefont {Scuseria}, \citenamefont {Constantin}, \citenamefont {Zhou},\
      and\ \citenamefont {Burke}}]{PBEsol}%
      \BibitemOpen
      \bibfield  {author} {\bibinfo {author} {\bibfnamefont {J.}~\bibnamefont
      {Perdew}}, \bibinfo {author} {\bibfnamefont {A.}~\bibnamefont {Ruzsinszky}},
      \bibinfo {author} {\bibfnamefont {G.}~\bibnamefont {Csonka}}, \bibinfo
      {author} {\bibfnamefont {O.}~\bibnamefont {Vydrov}}, \bibinfo {author}
      {\bibfnamefont {G.}~\bibnamefont {Scuseria}}, \bibinfo {author}
      {\bibfnamefont {L.}~\bibnamefont {Constantin}}, \bibinfo {author}
      {\bibfnamefont {X.}~\bibnamefont {Zhou}},\ and\ \bibinfo {author}
      {\bibfnamefont {K.}~\bibnamefont {Burke}},\ }\bibfield  {title}
      {{\selectlanguage {English}\bibinfo {title} {{Restoring the Density-Gradient
      Expansion for Exchange in Solids and Surfaces}}},\ }\href
      {https://doi.org/10.1103/physrevlett.100.136406} {\bibfield  {journal}
      {\bibinfo  {journal} {Phys. Rev. Lett.}\ }\textbf {\bibinfo {volume} {100}},\
      \bibinfo {pages} {136406} (\bibinfo {year} {2008})}\BibitemShut {NoStop}%
    \bibitem [{\citenamefont {Tadano}\ \emph {et~al.}(2014)\citenamefont {Tadano},
      \citenamefont {Gohda},\ and\ \citenamefont {Tsuneyuki}}]{Tadano2014}%
      \BibitemOpen
      \bibfield  {author} {\bibinfo {author} {\bibfnamefont {T.}~\bibnamefont
      {Tadano}}, \bibinfo {author} {\bibfnamefont {Y.}~\bibnamefont {Gohda}},\ and\
      \bibinfo {author} {\bibfnamefont {S.}~\bibnamefont {Tsuneyuki}},\ }\bibfield
      {title} {{\selectlanguage {English}\bibinfo {title} {{Anharmonic force
      constants extracted from first-principles molecular dynamics: applications to
      heat transfer simulations}}},\ }\href
      {https://doi.org/10.1088/0953-8984/26/22/225402} {\bibfield  {journal}
      {\bibinfo  {journal} {J. Phys: Condens. Matter}\ }\textbf {\bibinfo {volume}
      {26}},\ \bibinfo {pages} {225402} (\bibinfo {year} {2014})}\BibitemShut
      {NoStop}%
    \bibitem [{\citenamefont {Zhou}\ \emph {et~al.}(2014)\citenamefont {Zhou},
      \citenamefont {Nielson}, \citenamefont {Xia},\ and\ \citenamefont
      {Ozoliņš}}]{Zhou_PRL2014}%
      \BibitemOpen
      \bibfield  {author} {\bibinfo {author} {\bibfnamefont {F.}~\bibnamefont
      {Zhou}}, \bibinfo {author} {\bibfnamefont {W.}~\bibnamefont {Nielson}},
      \bibinfo {author} {\bibfnamefont {Y.}~\bibnamefont {Xia}},\ and\ \bibinfo
      {author} {\bibfnamefont {V.}~\bibnamefont {Ozoliņš}},\ }\bibfield  {title}
      {{\selectlanguage {English}\bibinfo {title} {{Lattice Anharmonicity and
      Thermal Conductivity from Compressive Sensing of First-Principles
      Calculations}}},\ }\href {https://doi.org/10.1103/physrevlett.113.185501}
      {\bibfield  {journal} {\bibinfo  {journal} {Phys. Rev. Lett.}\ }\textbf
      {\bibinfo {volume} {113}},\ \bibinfo {pages} {185501} (\bibinfo {year}
      {2014})}\BibitemShut {NoStop}%
    \bibitem [{\citenamefont {Zou}(2006)}]{adalasso}%
      \BibitemOpen
      \bibfield  {author} {\bibinfo {author} {\bibfnamefont {H.}~\bibnamefont
      {Zou}},\ }\bibfield  {title} {\bibinfo {title} {The adaptive lasso and its
      oracle properties},\ }\href {https://doi.org/10.1198/016214506000000735}
      {\bibfield  {journal} {\bibinfo  {journal} {J. Am. Stat. Assoc.}\ }\textbf
      {\bibinfo {volume} {101}},\ \bibinfo {pages} {1418} (\bibinfo {year}
      {2006})}\BibitemShut {NoStop}%
    \bibitem [{sup()}]{supplement}%
      \BibitemOpen
      \href@noop {} {}\bibinfo {note} {See Supplementary Material at [URL] for the
      detailed computational procedures, the temperature-dependent phonon
      dispersion curves of $\alpha$-\cpb calculated by SC1, QP[0], QP[S], and
      QP-NL, the lattice constant dependence of various anharmonic properties and
      \tc, and comparison of calculated and experimental linewidths of transverse
      acoustic phonons, which includes Refs.~\cite{Prandini_npjComputMater2018,
      Baroni_RMP2001, lassobook2015, Fransson_npjcomp2020, Wang_JPCM2010,
      Gonze_PRB1997,shengbte}}\BibitemShut {NoStop}%
    \bibitem [{\citenamefont {Rodová}\ \emph {et~al.}(2003)\citenamefont
      {Rodová}, \citenamefont {Brožek}, \citenamefont {Knížek},\ and\
      \citenamefont {Nitsch}}]{Rodova_JTAC2003}%
      \BibitemOpen
      \bibfield  {author} {\bibinfo {author} {\bibfnamefont {M.}~\bibnamefont
      {Rodová}}, \bibinfo {author} {\bibfnamefont {J.}~\bibnamefont {Brožek}},
      \bibinfo {author} {\bibfnamefont {K.}~\bibnamefont {Knížek}},\ and\
      \bibinfo {author} {\bibfnamefont {K.}~\bibnamefont {Nitsch}},\ }\bibfield
      {title} {\bibinfo {title} {{Phase transitions in ternary caesium lead
      bromide}},\ }\href {https://doi.org/10.1023/a:1022836800820} {\bibfield
      {journal} {\bibinfo  {journal} {J. Therm. Anal. Calorim.}\ }\textbf {\bibinfo
      {volume} {71}},\ \bibinfo {pages} {667} (\bibinfo {year} {2003})}\BibitemShut
      {NoStop}%
    \bibitem [{\citenamefont {Svirskas}\ \emph {et~al.}(2020)\citenamefont
      {Svirskas}, \citenamefont {Balči\={u}nas}, \citenamefont {Šim\.{e}nas},
      \citenamefont {Usevi\v{c}ius}, \citenamefont {Kinka}, \citenamefont
      {Veli\v{c}ka}, \citenamefont {Kubicki}, \citenamefont {Castillo},
      \citenamefont {Karabanov}, \citenamefont {Shvartsman}, \citenamefont
      {Soares}, \citenamefont {\v{S}ablinskas}, \citenamefont {Salak},
      \citenamefont {Lupascu},\ and\ \citenamefont {Banys}}]{Svirskas_JMCA}%
      \BibitemOpen
      \bibfield  {author} {\bibinfo {author} {\bibfnamefont {S.}~\bibnamefont
      {Svirskas}}, \bibinfo {author} {\bibfnamefont {S.}~\bibnamefont
      {Balči\={u}nas}}, \bibinfo {author} {\bibfnamefont {M.}~\bibnamefont
      {Šim\.{e}nas}}, \bibinfo {author} {\bibfnamefont {G.}~\bibnamefont
      {Usevi\v{c}ius}}, \bibinfo {author} {\bibfnamefont {M.}~\bibnamefont
      {Kinka}}, \bibinfo {author} {\bibfnamefont {M.}~\bibnamefont {Veli\v{c}ka}},
      \bibinfo {author} {\bibfnamefont {D.}~\bibnamefont {Kubicki}}, \bibinfo
      {author} {\bibfnamefont {M.~E.}\ \bibnamefont {Castillo}}, \bibinfo {author}
      {\bibfnamefont {A.}~\bibnamefont {Karabanov}}, \bibinfo {author}
      {\bibfnamefont {V.~V.}\ \bibnamefont {Shvartsman}}, \bibinfo {author}
      {\bibfnamefont {M.~d.~R.}\ \bibnamefont {Soares}}, \bibinfo {author}
      {\bibfnamefont {V.}~\bibnamefont {\v{S}ablinskas}}, \bibinfo {author}
      {\bibfnamefont {A.~N.}\ \bibnamefont {Salak}}, \bibinfo {author}
      {\bibfnamefont {D.~C.}\ \bibnamefont {Lupascu}},\ and\ \bibinfo {author}
      {\bibfnamefont {J.}~\bibnamefont {Banys}},\ }\bibfield  {title} {\bibinfo
      {title} {{Phase transitions, screening and dielectric response of
      CsPbBr$_3$}},\ }\href {https://doi.org/10.1039/d0ta04155f} {\bibfield
      {journal} {\bibinfo  {journal} {J. Mater. Chem. A}\ }\textbf {\bibinfo
      {volume} {8}},\ \bibinfo {pages} {14015} (\bibinfo {year}
      {2020})}\BibitemShut {NoStop}%
    \bibitem [{\citenamefont {Ehsan}\ \emph {et~al.}(2021)\citenamefont {Ehsan},
      \citenamefont {Arrigoni}, \citenamefont {Madsen}, \citenamefont {Blaha},\
      and\ \citenamefont {Tröster}}]{Ehsan_PRB2021}%
      \BibitemOpen
      \bibfield  {author} {\bibinfo {author} {\bibfnamefont {S.}~\bibnamefont
      {Ehsan}}, \bibinfo {author} {\bibfnamefont {M.}~\bibnamefont {Arrigoni}},
      \bibinfo {author} {\bibfnamefont {G.~K.~H.}\ \bibnamefont {Madsen}}, \bibinfo
      {author} {\bibfnamefont {P.}~\bibnamefont {Blaha}},\ and\ \bibinfo {author}
      {\bibfnamefont {A.}~\bibnamefont {Tröster}},\ }\bibfield  {title} {\bibinfo
      {title} {{First-principles self-consistent phonon approach to the study of
      the vibrational properties and structural phase transition of BaTiO$_3$}},\
      }\href {https://doi.org/10.1103/physrevb.103.094108} {\bibfield  {journal}
      {\bibinfo  {journal} {Phys. Rev. B}\ }\textbf {\bibinfo {volume} {103}},\
      \bibinfo {pages} {094108} (\bibinfo {year} {2021})}\BibitemShut {NoStop}%
    \bibitem [{\citenamefont {Tadano}\ and\ \citenamefont
      {Tsuneyuki}(2018{\natexlab{b}})}]{Tadano_PRL2018}%
      \BibitemOpen
      \bibfield  {author} {\bibinfo {author} {\bibfnamefont {T.}~\bibnamefont
      {Tadano}}\ and\ \bibinfo {author} {\bibfnamefont {S.}~\bibnamefont
      {Tsuneyuki}},\ }\bibfield  {title} {{\selectlanguage {English}\bibinfo
      {title} {{Quartic Anharmonicity of Rattlers and Its Effect on Lattice Thermal
      Conductivity of Clathrates from First Principles}}},\ }\href
      {https://doi.org/10.1103/physrevlett.120.105901} {\bibfield  {journal}
      {\bibinfo  {journal} {Phys. Rev. Lett.}\ }\textbf {\bibinfo {volume} {120}},\
      \bibinfo {pages} {105901} (\bibinfo {year} {2018}{\natexlab{b}})}\BibitemShut
      {NoStop}%
    \bibitem [{\citenamefont {Songvilay}\ \emph {et~al.}(2019)\citenamefont
      {Songvilay}, \citenamefont {Giles-Donovan}, \citenamefont {Bari},
      \citenamefont {Ye}, \citenamefont {Minns}, \citenamefont {Green},
      \citenamefont {Xu}, \citenamefont {Gehring}, \citenamefont {Schmalzl},
      \citenamefont {Ratcliff}, \citenamefont {Brown}, \citenamefont {Chernyshov},
      \citenamefont {Beek}, \citenamefont {Cochran},\ and\ \citenamefont
      {Stock}}]{Songvilay_PRM2019}%
      \BibitemOpen
      \bibfield  {author} {\bibinfo {author} {\bibfnamefont {M.}~\bibnamefont
      {Songvilay}}, \bibinfo {author} {\bibfnamefont {N.}~\bibnamefont
      {Giles-Donovan}}, \bibinfo {author} {\bibfnamefont {M.}~\bibnamefont {Bari}},
      \bibinfo {author} {\bibfnamefont {Z.-G.}\ \bibnamefont {Ye}}, \bibinfo
      {author} {\bibfnamefont {J.~L.}\ \bibnamefont {Minns}}, \bibinfo {author}
      {\bibfnamefont {M.~A.}\ \bibnamefont {Green}}, \bibinfo {author}
      {\bibfnamefont {G.}~\bibnamefont {Xu}}, \bibinfo {author} {\bibfnamefont
      {P.~M.}\ \bibnamefont {Gehring}}, \bibinfo {author} {\bibfnamefont
      {K.}~\bibnamefont {Schmalzl}}, \bibinfo {author} {\bibfnamefont {W.~D.}\
      \bibnamefont {Ratcliff}}, \bibinfo {author} {\bibfnamefont {C.~M.}\
      \bibnamefont {Brown}}, \bibinfo {author} {\bibfnamefont {D.}~\bibnamefont
      {Chernyshov}}, \bibinfo {author} {\bibfnamefont {W.~v.}\ \bibnamefont
      {Beek}}, \bibinfo {author} {\bibfnamefont {S.}~\bibnamefont {Cochran}},\ and\
      \bibinfo {author} {\bibfnamefont {C.}~\bibnamefont {Stock}},\ }\bibfield
      {title} {\bibinfo {title} {{Common acoustic phonon lifetimes in inorganic and
      hybrid lead halide perovskites}},\ }\href
      {https://doi.org/10.1103/physrevmaterials.3.093602} {\bibfield  {journal}
      {\bibinfo  {journal} {Phys. Rev. Mater.}\ }\textbf {\bibinfo {volume} {3}},\
      \bibinfo {pages} {093602} (\bibinfo {year} {2019})}\BibitemShut {NoStop}%
    \bibitem [{\citenamefont {Tripathi}\ and\ \citenamefont {Pathak}(4
      06)}]{Tripathi_Cimento1974}%
      \BibitemOpen
      \bibfield  {author} {\bibinfo {author} {\bibfnamefont {R.~S.}\ \bibnamefont
      {Tripathi}}\ and\ \bibinfo {author} {\bibfnamefont {K.~N.}\ \bibnamefont
      {Pathak}},\ }\bibfield  {title} {\bibinfo {title} {{Self-energy of phonons in
      an anharmonic crystal to O($\delta^4$)}},\ }\href
      {https://doi.org/10.1007/bf02737485} {\bibfield  {journal} {\bibinfo
      {journal} {Il Nuovo Cimento B}\ }\textbf {\bibinfo {volume} {21}},\ \bibinfo
      {pages} {289} (\bibinfo {year} {1974-06})}\BibitemShut {NoStop}%
    \bibitem [{\citenamefont {Feng}\ \emph {et~al.}(2017)\citenamefont {Feng},
      \citenamefont {Lindsay},\ and\ \citenamefont {Ruan}}]{Feng_PRB2017}%
      \BibitemOpen
      \bibfield  {author} {\bibinfo {author} {\bibfnamefont {T.}~\bibnamefont
      {Feng}}, \bibinfo {author} {\bibfnamefont {L.}~\bibnamefont {Lindsay}},\ and\
      \bibinfo {author} {\bibfnamefont {X.}~\bibnamefont {Ruan}},\ }\bibfield
      {title} {{\selectlanguage {English}\bibinfo {title} {{Four-phonon scattering
      significantly reduces intrinsic thermal conductivity of solids}}},\ }\href
      {https://doi.org/10.1103/physrevb.96.161201} {\bibfield  {journal} {\bibinfo
      {journal} {Phys. Rev. B}\ }\textbf {\bibinfo {volume} {96}},\ \bibinfo
      {pages} {72 } (\bibinfo {year} {2017})}\BibitemShut {NoStop}%
    \bibitem [{\citenamefont {Allen}\ and\ \citenamefont
      {Feldman}(1993)}]{Allen_PRB1993}%
      \BibitemOpen
      \bibfield  {author} {\bibinfo {author} {\bibfnamefont {P.~B.}\ \bibnamefont
      {Allen}}\ and\ \bibinfo {author} {\bibfnamefont {J.~L.}\ \bibnamefont
      {Feldman}},\ }\bibfield  {title} {{\selectlanguage {English}\bibinfo {title}
      {{Thermal conductivity of disordered harmonic solids}}},\ }\href
      {https://doi.org/10.1103/physrevb.48.12581} {\bibfield  {journal} {\bibinfo
      {journal} {Phys. Rev. B}\ }\textbf {\bibinfo {volume} {48}},\ \bibinfo
      {pages} {12581 } (\bibinfo {year} {1993})}\BibitemShut {NoStop}%
    \bibitem [{\citenamefont {Sun}\ and\ \citenamefont
      {Allen}(2010)}]{Sun_PRB2010}%
      \BibitemOpen
      \bibfield  {author} {\bibinfo {author} {\bibfnamefont {T.}~\bibnamefont
      {Sun}}\ and\ \bibinfo {author} {\bibfnamefont {P.~B.}\ \bibnamefont
      {Allen}},\ }\bibfield  {title} {\bibinfo {title} {{Lattice thermal
      conductivity: Computations and theory of the high-temperature breakdown of
      the phonon-gas model}},\ }\href {https://doi.org/10.1103/physrevb.82.224305}
      {\bibfield  {journal} {\bibinfo  {journal} {Phys. Rev. B}\ }\textbf {\bibinfo
      {volume} {82}},\ \bibinfo {pages} {224305} (\bibinfo {year}
      {2010})}\BibitemShut {NoStop}%
    \bibitem [{\citenamefont {Elbaz}\ \emph {et~al.}(2017)\citenamefont {Elbaz},
      \citenamefont {Ong}, \citenamefont {Doud}, \citenamefont {Kim}, \citenamefont
      {Paley}, \citenamefont {Roy},\ and\ \citenamefont
      {Malen}}]{Elbaz_NanoLetters2017}%
      \BibitemOpen
      \bibfield  {author} {\bibinfo {author} {\bibfnamefont {G.~A.}\ \bibnamefont
      {Elbaz}}, \bibinfo {author} {\bibfnamefont {W.-L.}\ \bibnamefont {Ong}},
      \bibinfo {author} {\bibfnamefont {E.~A.}\ \bibnamefont {Doud}}, \bibinfo
      {author} {\bibfnamefont {P.}~\bibnamefont {Kim}}, \bibinfo {author}
      {\bibfnamefont {D.~W.}\ \bibnamefont {Paley}}, \bibinfo {author}
      {\bibfnamefont {X.}~\bibnamefont {Roy}},\ and\ \bibinfo {author}
      {\bibfnamefont {J.~A.}\ \bibnamefont {Malen}},\ }\bibfield  {title} {\bibinfo
      {title} {{Phonon Speed, Not Scattering, Differentiates Thermal Transport in
      Lead Halide Perovskites}},\ }\href
      {https://doi.org/10.1021/acs.nanolett.7b02696} {\bibfield  {journal}
      {\bibinfo  {journal} {Nano Lett.}\ }\textbf {\bibinfo {volume} {17}},\
      \bibinfo {pages} {5734} (\bibinfo {year} {2017})}\BibitemShut {NoStop}%
    \bibitem [{\citenamefont {Wang}\ and\ \citenamefont
      {Lin}(2016)}]{Wang_AdvFuncMater2016}%
      \BibitemOpen
      \bibfield  {author} {\bibinfo {author} {\bibfnamefont {M.}~\bibnamefont
      {Wang}}\ and\ \bibinfo {author} {\bibfnamefont {S.}~\bibnamefont {Lin}},\
      }\bibfield  {title} {\bibinfo {title} {{Anisotropic and Ultralow Phonon
      Thermal Transport in Organic–Inorganic Hybrid Perovskites: Atomistic
      Insights into Solar Cell Thermal Management and Thermoelectric Energy
      Conversion Efficiency}},\ }\href {https://doi.org/10.1002/adfm.201600284}
      {\bibfield  {journal} {\bibinfo  {journal} {Adv. Funct. Mater.}\ }\textbf
      {\bibinfo {volume} {26}},\ \bibinfo {pages} {5297} (\bibinfo {year}
      {2016})}\BibitemShut {NoStop}%
    \bibitem [{\citenamefont {Lee}\ \emph {et~al.}(2017)\citenamefont {Lee},
      \citenamefont {Li}, \citenamefont {Wong}, \citenamefont {Zhang},
      \citenamefont {Lai}, \citenamefont {Yu}, \citenamefont {Kong}, \citenamefont
      {Lin}, \citenamefont {Urban}, \citenamefont {Grossman},\ and\ \citenamefont
      {Yang}}]{Lee_PNAS2017}%
      \BibitemOpen
      \bibfield  {author} {\bibinfo {author} {\bibfnamefont {W.}~\bibnamefont
      {Lee}}, \bibinfo {author} {\bibfnamefont {H.}~\bibnamefont {Li}}, \bibinfo
      {author} {\bibfnamefont {A.~B.}\ \bibnamefont {Wong}}, \bibinfo {author}
      {\bibfnamefont {D.}~\bibnamefont {Zhang}}, \bibinfo {author} {\bibfnamefont
      {M.}~\bibnamefont {Lai}}, \bibinfo {author} {\bibfnamefont {Y.}~\bibnamefont
      {Yu}}, \bibinfo {author} {\bibfnamefont {Q.}~\bibnamefont {Kong}}, \bibinfo
      {author} {\bibfnamefont {E.}~\bibnamefont {Lin}}, \bibinfo {author}
      {\bibfnamefont {J.~J.}\ \bibnamefont {Urban}}, \bibinfo {author}
      {\bibfnamefont {J.~C.}\ \bibnamefont {Grossman}},\ and\ \bibinfo {author}
      {\bibfnamefont {P.}~\bibnamefont {Yang}},\ }\bibfield  {title}
      {{\selectlanguage {English}\bibinfo {title} {{Ultralow thermal conductivity
      in all-inorganic halide perovskites}}},\ }\href
      {https://doi.org/10.1073/pnas.1711744114} {\bibfield  {journal} {\bibinfo
      {journal} {Proc. Natl. Acad. Sci.}\ }\textbf {\bibinfo {volume} {114}},\
      \bibinfo {pages} {8693} (\bibinfo {year} {2017})}\BibitemShut {NoStop}%
    \bibitem [{\citenamefont {Prandini}\ \emph {et~al.}(2018)\citenamefont
      {Prandini}, \citenamefont {Marrazzo}, \citenamefont {Castelli}, \citenamefont
      {Mounet},\ and\ \citenamefont {Marzari}}]{Prandini_npjComputMater2018}%
      \BibitemOpen
      \bibfield  {author} {\bibinfo {author} {\bibfnamefont {G.}~\bibnamefont
      {Prandini}}, \bibinfo {author} {\bibfnamefont {A.}~\bibnamefont {Marrazzo}},
      \bibinfo {author} {\bibfnamefont {I.~E.}\ \bibnamefont {Castelli}}, \bibinfo
      {author} {\bibfnamefont {N.}~\bibnamefont {Mounet}},\ and\ \bibinfo {author}
      {\bibfnamefont {N.}~\bibnamefont {Marzari}},\ }\bibfield  {title} {\bibinfo
      {title} {{Precision and efficiency in solid-state pseudopotential
      calculations}},\ }\href {https://doi.org/10.1038/s41524-018-0127-2}
      {\bibfield  {journal} {\bibinfo  {journal} {npj Comput. Mater.}\ }\textbf
      {\bibinfo {volume} {4}},\ \bibinfo {pages} {72} (\bibinfo {year}
      {2018})}\BibitemShut {NoStop}%
    \bibitem [{\citenamefont {Baroni}\ \emph {et~al.}(2001)\citenamefont {Baroni},
      \citenamefont {Gironcoli}, \citenamefont {Corso},\ and\ \citenamefont
      {Giannozzi}}]{Baroni_RMP2001}%
      \BibitemOpen
      \bibfield  {author} {\bibinfo {author} {\bibfnamefont {S.}~\bibnamefont
      {Baroni}}, \bibinfo {author} {\bibfnamefont {S.~d.}\ \bibnamefont
      {Gironcoli}}, \bibinfo {author} {\bibfnamefont {A.~D.}\ \bibnamefont
      {Corso}},\ and\ \bibinfo {author} {\bibfnamefont {P.}~\bibnamefont
      {Giannozzi}},\ }\bibfield  {title} {\bibinfo {title} {{Phonons and related
      crystal properties from density-functional perturbation theory}},\ }\href
      {https://doi.org/10.1103/revmodphys.73.515} {\bibfield  {journal} {\bibinfo
      {journal} {Rev. Mod. Phys.}\ }\textbf {\bibinfo {volume} {73}},\ \bibinfo
      {pages} {515} (\bibinfo {year} {2001})}\BibitemShut {NoStop}%
    \bibitem [{\citenamefont {Hastie}\ \emph {et~al.}(2015)\citenamefont {Hastie},
      \citenamefont {Tibshirani},\ and\ \citenamefont
      {Wainwright}}]{lassobook2015}%
      \BibitemOpen
      \bibfield  {author} {\bibinfo {author} {\bibfnamefont {T.}~\bibnamefont
      {Hastie}}, \bibinfo {author} {\bibfnamefont {R.}~\bibnamefont {Tibshirani}},\
      and\ \bibinfo {author} {\bibfnamefont {M.}~\bibnamefont {Wainwright}},\
      }\href@noop {} {\emph {\bibinfo {title} {{Statistical Learning with Sparsity:
      The Lasso and Generalizations}}}},\ Chapman \& Hall/CRC\ (\bibinfo
      {publisher} {Chapman \& Hall/CRC},\ \bibinfo {year} {2015})\BibitemShut
      {NoStop}%
    \bibitem [{\citenamefont {Fransson}\ \emph {et~al.}(2020)\citenamefont
      {Fransson}, \citenamefont {Eriksson},\ and\ \citenamefont
      {Erhart}}]{Fransson_npjcomp2020}%
      \BibitemOpen
      \bibfield  {author} {\bibinfo {author} {\bibfnamefont {E.}~\bibnamefont
      {Fransson}}, \bibinfo {author} {\bibfnamefont {F.}~\bibnamefont {Eriksson}},\
      and\ \bibinfo {author} {\bibfnamefont {P.}~\bibnamefont {Erhart}},\
      }\bibfield  {title} {\bibinfo {title} {{Efficient construction of linear
      models in materials modeling and applications to force constant
      expansions}},\ }\href {https://doi.org/10.1038/s41524-020-00404-5} {\bibfield
       {journal} {\bibinfo  {journal} {npj Comput. Mater.}\ }\textbf {\bibinfo
      {volume} {6}},\ \bibinfo {pages} {135} (\bibinfo {year} {2020})}\BibitemShut
      {NoStop}%
    \bibitem [{\citenamefont {Wang}\ \emph {et~al.}(2010)\citenamefont {Wang},
      \citenamefont {Wang}, \citenamefont {Wang}, \citenamefont {Mei},
      \citenamefont {Shang}, \citenamefont {Chen},\ and\ \citenamefont
      {Liu}}]{Wang_JPCM2010}%
      \BibitemOpen
      \bibfield  {author} {\bibinfo {author} {\bibfnamefont {Y.}~\bibnamefont
      {Wang}}, \bibinfo {author} {\bibfnamefont {J.~J.}\ \bibnamefont {Wang}},
      \bibinfo {author} {\bibfnamefont {W.~Y.}\ \bibnamefont {Wang}}, \bibinfo
      {author} {\bibfnamefont {Z.~G.}\ \bibnamefont {Mei}}, \bibinfo {author}
      {\bibfnamefont {S.~L.}\ \bibnamefont {Shang}}, \bibinfo {author}
      {\bibfnamefont {L.~Q.}\ \bibnamefont {Chen}},\ and\ \bibinfo {author}
      {\bibfnamefont {Z.~K.}\ \bibnamefont {Liu}},\ }\bibfield  {title} {\bibinfo
      {title} {{A mixed-space approach to first-principles calculations of phonon
      frequencies for polar materials}},\ }\href
      {https://doi.org/10.1088/0953-8984/22/20/202201} {\bibfield  {journal}
      {\bibinfo  {journal} {J. Phys.: Condens. Matter}\ }\textbf {\bibinfo {volume}
      {22}},\ \bibinfo {pages} {202201} (\bibinfo {year} {2010})}\BibitemShut
      {NoStop}%
    \bibitem [{\citenamefont {Gonze}\ and\ \citenamefont
      {Lee}(1997)}]{Gonze_PRB1997}%
      \BibitemOpen
      \bibfield  {author} {\bibinfo {author} {\bibfnamefont {X.}~\bibnamefont
      {Gonze}}\ and\ \bibinfo {author} {\bibfnamefont {C.}~\bibnamefont {Lee}},\
      }\bibfield  {title} {\bibinfo {title} {{Dynamical matrices, Born effective
      charges, dielectric permittivity tensors, and interatomic force constants
      from density-functional perturbation theory}},\ }\href
      {https://doi.org/10.1103/physrevb.55.10355} {\bibfield  {journal} {\bibinfo
      {journal} {Phys. Rev. B}\ }\textbf {\bibinfo {volume} {55}},\ \bibinfo
      {pages} {10355} (\bibinfo {year} {1997})}\BibitemShut {NoStop}%
    \bibitem [{\citenamefont {Li}\ \emph {et~al.}(2014)\citenamefont {Li},
      \citenamefont {Carrete}, \citenamefont {Katcho},\ and\ \citenamefont
      {Mingo}}]{shengbte}%
      \BibitemOpen
      \bibfield  {author} {\bibinfo {author} {\bibfnamefont {W.}~\bibnamefont
      {Li}}, \bibinfo {author} {\bibfnamefont {J.}~\bibnamefont {Carrete}},
      \bibinfo {author} {\bibfnamefont {N.~A.}\ \bibnamefont {Katcho}},\ and\
      \bibinfo {author} {\bibfnamefont {N.}~\bibnamefont {Mingo}},\ }\bibfield
      {title} {\bibinfo {title} {{ShengBTE: A solver of the Boltzmann transport
      equation for phonons}},\ }\href {https://doi.org/10.1016/j.cpc.2014.02.015}
      {\bibfield  {journal} {\bibinfo  {journal} {Comput. Phys. Commun.}\ }\textbf
      {\bibinfo {volume} {185}},\ \bibinfo {pages} {1 } (\bibinfo {year}
      {2014})}\BibitemShut {NoStop}%
    \end{thebibliography}
\end{document}